\documentclass[manuscript]{aastex}

\slugcomment{Not to appear in Nonlearned J., 45.}

\shorttitle{A family of zero-velocity curves}
\shortauthors{Roman et al.}

\begin{document}


\title{A family of zero-velocity curves in the restricted three-body problem}


\author{R. Roman$\dagger$\altaffilmark{1} and I. Sz{\"u}cs-Csillik\altaffilmark{1}}
\affil{Astronomical Institute of Romanian Academy, Astronomical Observatory Cluj-Napoca, Str. Ciresilor No. 19, RO-400487 Cluj-Napoca, Romania}

\email{iharka@gmail.com}

\begin{abstract}
The equilibrium points and the curves of zero-velocity
(Roche varieties) are analyzed in the frame of the regularized circular restricted three-body problem. The coordinate transformation is done with Levi-Civita generalized method, using polynomial functions of $n$
degree. In the parametric plane, five families of equilibrium
points are identified: $L_i^1, \;, L_i^2, ..., \;L_i^n$, $i\in\{
1,2,...,5  \}, n \in \mathbb{N}^*$. 
These families of points correspond to the five
equilibrium points in the physical plane $L_1, \;L_2, ..., \;L_5$.
The zero-velocity curves from the physical plane are transformed
in Roche varieties in the parametric plane. The properties of
these varieties are analyzed and the Roche varieties for $n\in\{
1,2,...,6  \}$ are plotted. The equation of the asymptotic variety
is obtained and its shape is analyzed. The slope of the Roche
variety in $L_1^1$ point is obtained. For $n=1$ the slope obtained
by Plavec and Kratochvil (1964) in the physical plane was found.
\end{abstract}


\section{Introduction}
The generalized Levi-Civita regularization method was described in
a previous article \citep{RR2014} and applied in the
regularization of the equations of motion of the test particle in
the restricted three-body problem. Briefly, this method uses a
family of harmonic and conjugate polynomials of $n$ degree, in
order to realize the coordinate transformation.  After a
convenient time transformation, a system of differential equations of motion
without singularities is obtained.

For $n=2$ the Levi-Civita regularization method is retrieved \citep{Levi1906}.

The generalized Levi-Civita regularization method preserves the
structure of two important geometrical properties of Levi-Civita
transformation concerning the radii and polar angles in the
physical and parametric plane \citep{RR2014}.

As Levi-Civita generalized geometrical transformation transforms
each point from the physical plane in one or more points in the
parametric plane, we can ask how certain special points of the
physical plane are transformed in the parametric plane. A
particular importance can have the answer to the question: how are
the points of zero velocity curves transformed from the physical
to the parametric plane? This answer gives us a better
understanding of the parametric plane, and of the trajectories
around the singularity points.

In the following we will appoint \textsl{Roche varieties}, the set of points in the parametric plane, which are obtained by the transformation of points of zero velocity curves of the physical plane by Levi-Civita generalized method.

In 1966, Szebehely and Pierce represented the curves of zero velocity in the parametric plane, using Levi-Civita, Thiele-Burrau and Lema\^{i}tre regularized methods \citep{Sze1966}.

In the following we shall plot and analyze the Roche varieties, when the generalized Levi-Civita regularization method is used, focusing on what the generalization produces in the curves' topology.

\section{Zero velocity curves in the parametric plane}

In the circular restricted three-body problem, the canonical equations of motion of the test particle can be written using the generalized coordinates $q_1, \; q_2$ and generalized momenta $p_1, \; p_2$, in the form (see \citet{RR2014}):
\begin{eqnarray}\label{eq1-3}
\frac{d q_1}{dt} &=& p_1+q_2, \\
\frac{d q_2}{dt} &=& p_2-q_1, \\
\frac{d p_1}{dt} &=& p_2 - \frac{q}{1+q}-\frac{1}{1+q}\cdot
\frac{q_1}{r_1^3} - \frac{q}{1+q} \cdot \frac{q_1-1}{r_2^3},  \\
\frac{d p_2}{dt} &=& -p_1 -\frac{1}{1+q}\cdot
\frac{q_2}{r_1^3} - \frac{q}{1+q} \cdot\frac{q_2}{r_2^3},
\end{eqnarray}
where
\begin{equation}
r_1=\sqrt{q_1^2+q_2^2},\;\;\;\;r_2=\sqrt{(q_1-1)^2+q_2^2}.
\end{equation}
Here $q=\frac{m_2}{m_1}$, with $m_1$ and $m_2$ the masses of the components $S_1$ and $S_2$ of the binary system, $m_1>m_2$, the comoving coordinate system having the origin in the center of the more massive star ($S_1$), and the motion of the test particle being considered in the orbital plane. Here $r_1$ and $r_2$ are the distances of the test particle to the components of the binary system.
The potential function is (see \cite{RR2014}):
\begin{equation}
\psi(q_1,q_2)=\frac{1}{2}\left[\left(q_1-\frac{q}{1+q}   \right)^2+q_2^2+\frac{2}{(1+q)r_1}  +\frac{2q}{(1+q)r_2} \right].
\end{equation}
In order to obtain the equations of motion in the physical plane, let us differentiate with respect to time the equations (1) and (2). Then, from eqs. (3) and (4) we have:
\begin{equation}
\frac{d^2q_1}{dt^2}-2\frac{dq_2}{dt}=q_1-\frac{q}{1+q}-\frac{q_1}{(1+q)r_1^3}-\frac{q(q_1-1)}{(1+q)r_2^3},
\end{equation}
\begin{equation}
\frac{d^2q_2}{dt^2}+2\frac{dq_1}{dt}=q_2-\frac{q_2}{(1+q)r_1^3}-\frac{q\:q_2}{(1+q)r_2^3}.
\end{equation}
The equipotential curves in physical plane are obtained from $\psi(q_1,q_2)\;=\;$const., namely:
\begin{equation}
q_1^2+q_2^2-2\frac{qq_1}{1+q}+\frac{2}{(1+q)r_1}  +\frac{2q}{(1+q)r_2} \;=\;C,
\end{equation}
$C$ being the Jacobi constant.
The equations of motion (7)-(8) have singularities in terms $\frac{1}{r_1}$ and $\frac{1}{r_2}$. These singularities can be eliminate by regularization.

In the generalized Levi-Civita regularization method, one makes the transformation of coordinates (see \cite{RR2014}):
\begin{eqnarray}
q_1=\Re\left[ (Q_1+iQ_2)^n  \right]\;,\;\;\;n\in \mathbb{N}^*,\\
q_2=\Im\left[ (Q_1+iQ_2)^n  \right]\;,\;\;\;n\in \mathbb{N}^*.
\end{eqnarray}
Using eqs. (10)-(11), the physical plane ($q_1, S_1, q_2$) is transformed in the parametric plane ($Q_1, S_1, Q_2$). From these equations one can see that, to the origin of the coordinate system in the physical plane it corresponds the origin of the coordinate system in the parametric plane, because from equations $q_1=0,\; q_2=0$ it results only $Q_1=0,\; Q_2=0$.

But to the point $S_2(1,0)$ from the physical plane it correspond $n$ points $S_2$ in the parametric plane, because the system (10)-(11), with $q_1=1$ and $ q_2=0$ has the solution:
$$ Q_1=\cos\frac{2k\pi}{n}, \;\;  Q_2=\sin\frac{2k\pi}{n},\;\;k\in \{ 0,1,...,n-1 \},\;\;n\in\mathbb{N}^*\;  $$
(see Table 2, in \cite{RR2014}).

These $n$ points are located into the vertices of a regular
$n$-sided polygon, having the radius equal to 1 and one of the
vertices situated into the point of coordinates $(1,0)$.

From eqs. (10)-(11) it results that the curves of zero velocity given by eq. (9) are transformed in curves given by equation:
\begin{equation}
\left( Q_1^2+Q_2^2 \right)^n-2\frac{q\;\Re\left[ (Q_1+iQ_2)^n  \right]}{1+q}+\frac{2}{(1+q)r_1}+\frac{2q}{(1+q)r_2}\;=\;C,
\end{equation}
where
$$ r_1=\sqrt{\left( Q_1^2+Q_2^2 \right)^n},\;\;r_2=\sqrt{\left( Q_1^2+Q_2^2 \right)^n-2\Re\left[ (Q_1+iQ_2)^n \right]+1}.   $$
Szebehely (1967) write that the curves given by eq. (12) are curves of zero velocity, and the curves given by eq. (9) are both curves of zero velocity and equipotentials. It results that the Roche varieties are curves of zero velocity.

\textbf{Theorem 1.}
\textit{To a point $A(a,b)$ with $a,b\in \mathbb{R}$,  from the physical plane, it correspond $n$ points in the parametric plane, situated in vertices of an $n$-sided regular polygon having the radius $\sqrt[2n]{a^2+b^2}$.}\\
\textit{Proof}:\\
From eqs. (10)-(11) one obtain
\begin{eqnarray}
a=\Re\left[ (Q_1+iQ_2)^n  \right],\;\;\;n\in \mathbb{N}^*,\\
b=\Im\left[ (Q_1+iQ_2)^n  \right],\;\;\;n\in \mathbb{N}^*.
\end{eqnarray}
Multiplying the second equation with $i=\sqrt{-1}$ and gathering the left hands and the right hands it results (see \cite{Cara2001}):
\begin{equation}
a+ib=(Q_1+iQ_2)^n
\end{equation}
or $$\sqrt{a^2+b^2}\left(\cos T^*+i\;\sin T^*   \right)=(Q_1+iQ_2)^n, \; n\in \mathbb{N}^*, \tan T^*=\frac{b}{a},$$
and then:
\begin{eqnarray}
Q_1=\sqrt[2n]{a^2+b^2}\;\cos \frac{T^*+2k\pi}{n},\;\;k\in\{0,1,...,n-1  \},  \\
Q_2=\sqrt[2n]{a^2+b^2}\;\sin \frac{T^*+2k\pi}{n},\;\;k\in\{0,1,...,n-1  \}.
\end{eqnarray}
The eqs. (16)-(17) show us that to the point $A(a,b)$ from the physical plane it correspond $n$ points $(Q_1, Q_2)$ in the parametric plane, located in the vertices of an $n$-sided regular polygon having the radius $\sqrt[2n]{a^2+b^2}$, but to the origin of the coordinate system in the physical plane ($a=b=0$) it corresponds only one point in the parametric plane, located in the origin of the coordinate system (the circle with radius  $\sqrt[2n]{a^2+b^2}$ becomes a point).

\textbf{Corollary 1.}
\textit{To the triangular equilibrium point $L_4(\frac{1}{2},\frac{\sqrt{3}}{2})$ from the physical plane, it correspond $n$ points in the parametric plane, situated in vertices of an $n$-sided regular polygon having the radius 1.}

\textit{Proof}: $\sqrt[2n]{a^2+b^2}=1$.

Similar to triangular equilibrium point $L_5(\frac{1}{2},-\frac{\sqrt{3}}{2})$.

\textbf{Corollary 2.}
\textit{To the equilibrium point $L_1({q_1}_{L_1}, 0)$ from the physical plane, ${q_1}_{L_1} \in (0,1)$,  it correspond $n$ points in the parametric plane, situated in vertices of an $n$-sided regular polygon having the radius $\sqrt[n]{{q_1}_{L_1}}$, one of these vertices being located on the $S_1Q_1$- axis}.

\textit{Proof}: From Theorem 1 for $a={q_1}_{L_1}\in (0,1)$ and $b=0$ it results $T^*=0$ and
$Q_1=\sqrt[n]{{q_1}_{L_1}}\;\cos \frac{2k\pi}{n}$, $Q_2=\sqrt[n]{{q_1}_{L_1}}\;\sin \frac{2k\pi}{n}$, $k\in\{ 0,1,...,n-1 \}$. For $k=0$ one have $Q_1=\sqrt[n]{{q_1}_{L_1}}$, $Q_2=0$, so the point $L_1^1(\sqrt[n]{{q_1}_{L_1}},\;0)$ is situated on the $S_1Q_1$ axis, the other $n-1$ points being located in the vertices of the $n$-sided regular polygon having the radius $\sqrt[n]{{q_1}_{L_1}}$.

\textbf{Corollary 3.}
\textit{To the equilibrium point $L_2({q_1}_{L_2}, 0)$ from the physical plane, ${q_1}_{L_2} > 1$,  it correspond $n$ points in the parametric plane, situated in vertices of an $n$-sided regular polygon having the radius $\sqrt[n]{{q_1}_{L_2}}$, one of these vertices being located on the $S_1Q_1$ - axis}.

\textit{Proof}: Similar with corollary 2, because from  ${q_1}_{L_2}>1$ one obtain $|{q_1}_{L_2}|={q_1}_{L_2}$.

\textbf{Corollary 4.}
\textit{To the equilibrium point $L_3({q_1}_{L_3}, 0)$ from the physical plane, ${q_1}_{L_3} < 0$,  it correspond $n$ points in the parametric plane, situated in vertices of an $n$-sided regular polygon having the radius $\sqrt[n]{{q_1}_{L_3}}$.\\
If $n=2p+1$, $p\in \mathbb{N}$, then one of these vertices is located on the $S_1Q_1$ - axis}.\\
If $n=2p$, $p\in \mathbb{N}$, $p$ \textit{odd number, then one of the vertices is located on the $S_1Q_2$ axis.}\\
If $n=4p$, $p\in \mathbb{N}$, \textit{then all the vertices of the polygon can be situated in the outside of the coordinate axis. }

\textit{Proof}: From Theorem 1, if $b=0$ it results $T^*=0$.

If $n=2p+1$, from eq. (15), with $a={q_1}_{L_3}<0$ we obtain:
$$Q_1+iQ_2=-\sqrt[2p+1]{|{q_1}_{L_3}|}(\cos \frac{2k\pi}{n}+i\sin\frac{2k\pi}{n}).$$
For $k=0$ we obtain a vertex on the $S_1Q_1$ axis.

If $n=2p$, $p$ odd, having in view that $\sqrt[2p]{{q_1}_{L_3}}=i\sqrt[2p]{|{q_1}_{L_3}|}$, we obtain:
$$Q_1=-\sqrt[2p]{|{q_1}_{L_3}|}\;\sin\frac{2k\pi}{n},\;\;Q_2=\sqrt[2p]{|{q_1}_{L_3}|}\;\cos\frac{2k\pi}{n},\;\;n\in\mathbb{N}^*,
\;\;k\in \{ 0,1,...,n-1  \},$$ so one vertex of the regular polygon is located on the $S_1Q_2$ axis (for $k=0$).

If $n=4p$, from eq. (15) we have: $a+bi=(Q_1+iQ_2)^{4p}.$ Considering $a={q_1}_{L_3}<0$, and $b=0$ we obtain:
$$-|{q_1}_{L_3}|=\left[Q_1^4+Q_2^4-6Q_1^2Q_2^2+4iQ_1Q_2(Q_1^2-Q_2^2)   \right]^p.$$
For $p=1$ it results:
$L_3^1\left( \frac{\sqrt[4]{|{q_1}_{L_3}|}}{2}, \frac{\sqrt[4]{|{q_1}_{L_3}|}}{2} \right)$; $L_3^2\left(- \frac{\sqrt[4]{|{q_1}_{L_3}|}}{2}, \frac{\sqrt[4]{|{q_1}_{L_3}|}}{2} \right)$; $L_3^3\left( -\frac{\sqrt[4]{|{q_1}_{L_3}|}}{2}, -\frac{\sqrt[4]{|{q_1}_{L_3}|}}{2} \right)$; $L_3^4\left( \frac{\sqrt[4]{|{q_1}_{L_3}|}}{2}, -\frac{\sqrt[4]{|{q_1}_{L_3}|}}{2} \right)$, so if $n=4p$, $p\in \mathbb{N}^*$, then all the vertices of the polygon can be situated in the outside of the coordinate axis.

\textit{Remarks 1}:
\begin{enumerate}
    \item In the physical plane there are 5  equilibrium points: $L_1$, $L_2$, $L_3$, $L_4$, $L_5$. In the parametric plane there are 5 families of equilibrium points, each family having $n$ members, denoted: $L_1^1, L_1^2,..., L_1^n$; $\;L_2^1, L_2^2,..., L_2^n$; ... , $\;L_5^1, L_5^2,..., L_5^n$.
    \item In the physical plane there aren't equilibrium points on the $S_1q_2$ axis, while in the parametric plane there are equilibrium points on $S_1Q_2$ axis, points from the family of $L_3$ if $n=2p$, $p$ impair; we can have also points from the family of $L_1$ and $L_2$ if $\frac{2k\pi}{n}$ is an impair multiple of $\frac{\pi}{2}$.
    \item In the physical plane one speak about "collinear" and "triangular equilibrium points". In the parametric plane one have to speak about "polygonal equilibrium points".
\end{enumerate}

An interesting analogy (from theoretical point of view) can be established between the equations which give the coordinates of the equilibrium points in the physical and parametric plane. This will be presented in the following.

It is well known that the coordinates of the equilibrium points in the physical plane are given by the equations \citep{Sze1967,RR2003}:
\begin{eqnarray}
&q_1&\left(1-\frac{1}{(1+q)r_1^3} -\frac{q}{(1+q)r_2^3}  \right)+\frac{q}{1+q}\left( \frac{1}{r_2^3}-1  \right)=0, \\
&q_2&\left(1-\frac{1}{(1+q)r_1^3} -\frac{q}{(1+q)r_2^3}  \right)=0,
\end{eqnarray}
where $r_1=\sqrt{q_1^2+q_2^2},\;\;\;\;r_2=\sqrt{(q_1-1)^2+q_2^2}.$

In order to obtain the positions of double points in the parametric plane, we have to solve the equations:
$$\frac{\partial\Psi(Q_1,Q_2)}{\partial Q_1}=0=\frac{\partial\Psi(Q_1,Q_2)}{\partial Q_2}, $$
where:
\begin{equation}
\Psi(Q_1,Q_2)=\frac{1}{2}\left\{(Q_1^2+Q_2^2)^n-2\frac{q\;\Re[(Q_1+iQ_2)^n]}{1+q} +\frac{2}{(1+q)r_1}+\frac{2q}{(1+q)r_2}  \right\},
\end{equation}
with
\begin{equation}
 r_1=\sqrt{\left( Q_1^2+Q_2^2 \right)^n},\;\;r_2=\sqrt{\left( Q_1^2+Q_2^2 \right)^n-2\Re\left[ (Q_1+iQ_2)^n \right]+1}.
\end{equation}
So, the equilibrium points in the parametric plane are obtained from the following equations:
\begin{eqnarray}
Q_1(Q_1^2+Q_2^2)^{n-1}\left[1-\frac{1}{(1+q)r_1^3}-\frac{q}{(1+q)r_2^3}   \right]+\frac{q}{1+q}\Re[(Q_1+iQ_2)^{n-1}]     \left( \frac{1}{r_2^3}-1  \right)=0,  \\
Q_2(Q_1^2+Q_2^2)^{n-1}\left[1-\frac{1}{(1+q)r_1^3}-\frac{q}{(1+q)r_2^3}   \right]-\frac{q}{1+q}\Im[(Q_1+iQ_2)^{n-1}]     \left( \frac{1}{r_2^3}-1  \right)=0.
\end{eqnarray}
For $n=1$ the equations (22)-(23) coincides with equations (18)-(19), which is normal, because if $n=1$ the parametric plane coincides with the physical plane.

Selecting a numerical value for $n$, and solving the eqs. (22)-(23) we obtain the coordinates of the equilibrium points from the five family and then the corresponding Jacobi constants. Then, using eq. (12) we can obtain the Roche varieties (curves of zero velocity) in the parametric plane.
In Figures 1, 2,...,6 we plotted these curves for $n\in \{1, 2,..., 6\}$.

\begin{figure}
\begin{center}
  \includegraphics[height=0.3\textheight]{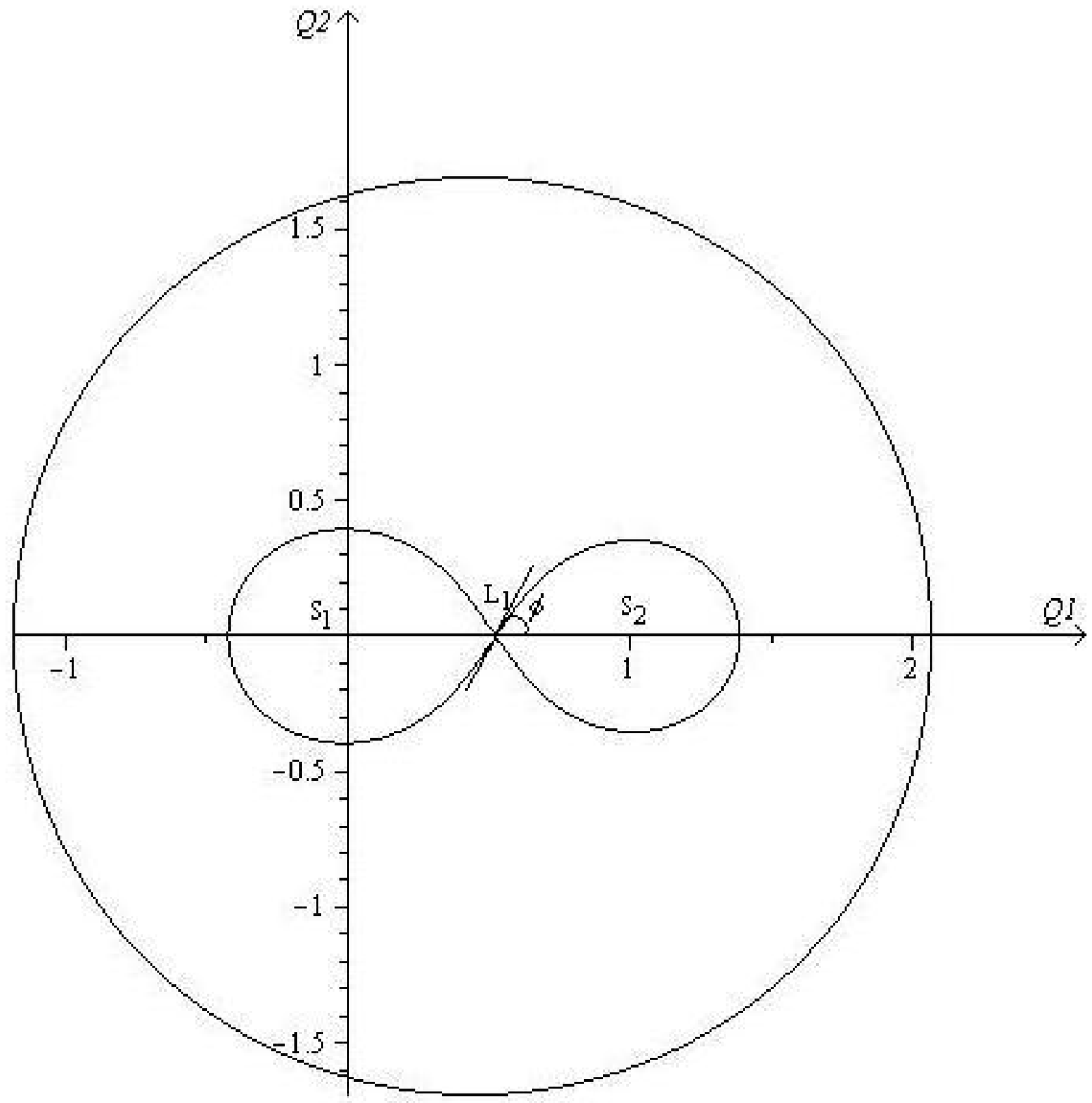}
    \includegraphics[height=0.3\textheight]{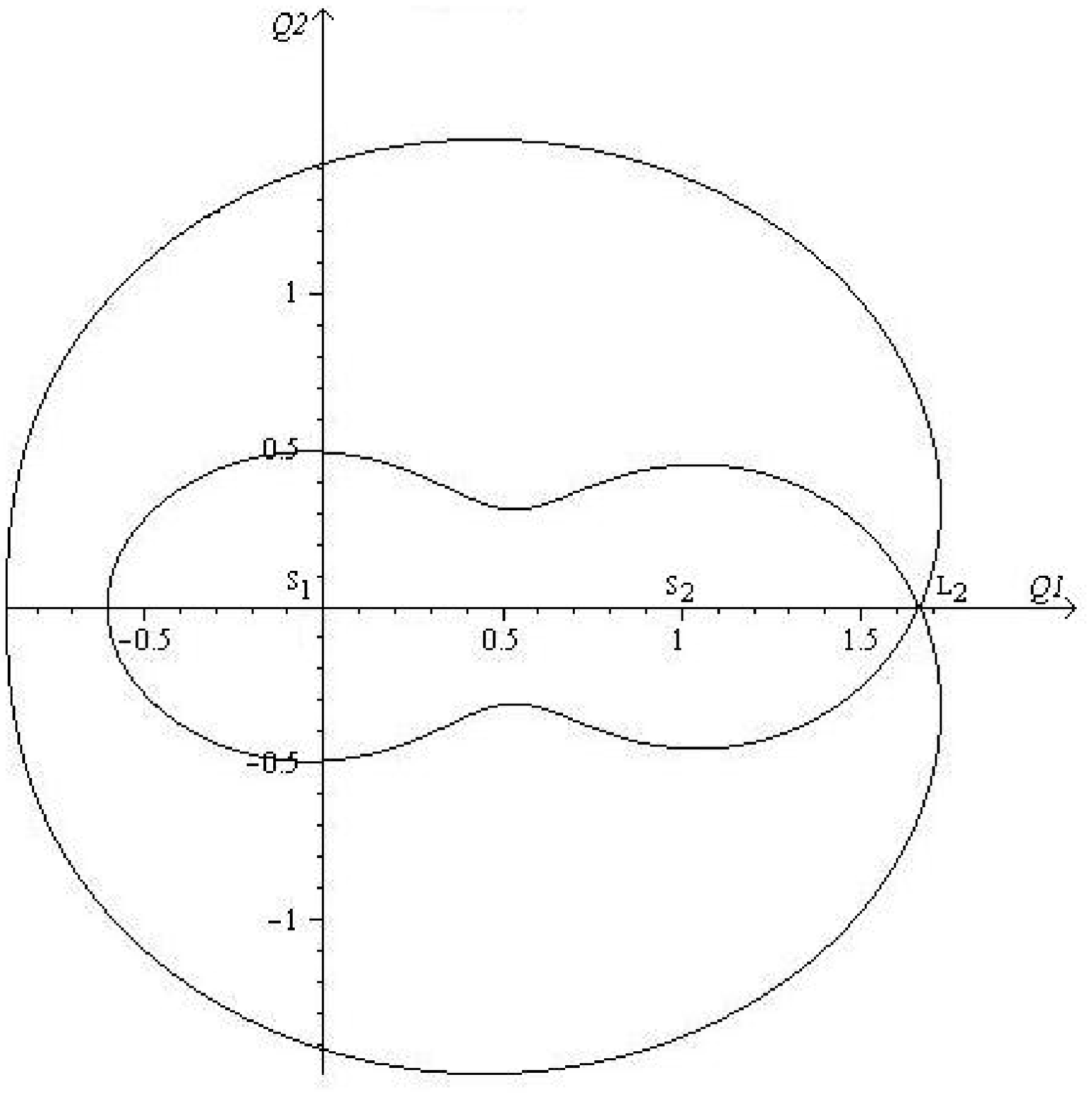}
    \includegraphics[height=0.3\textheight]{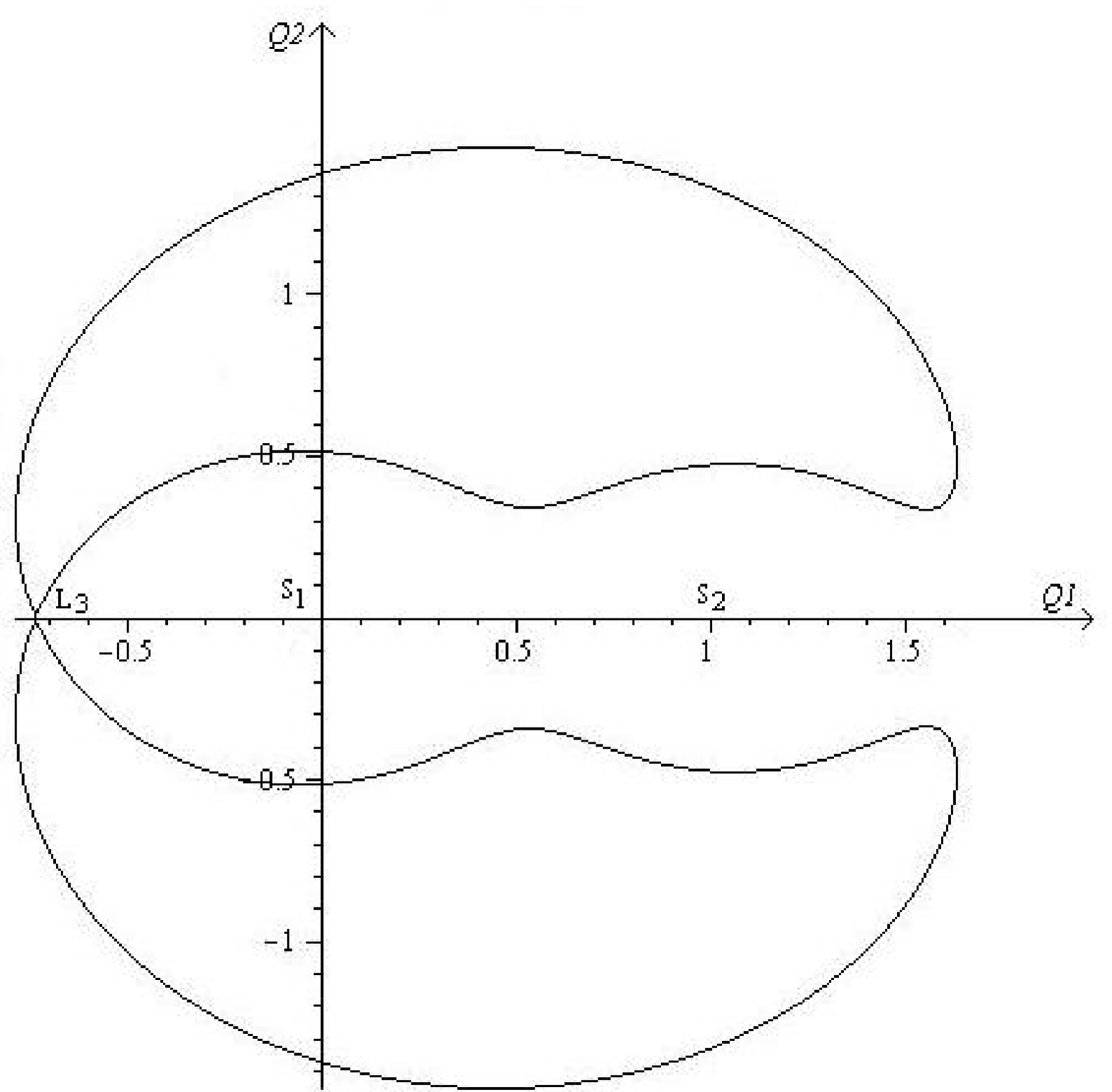}
    \includegraphics[height=0.3\textheight]{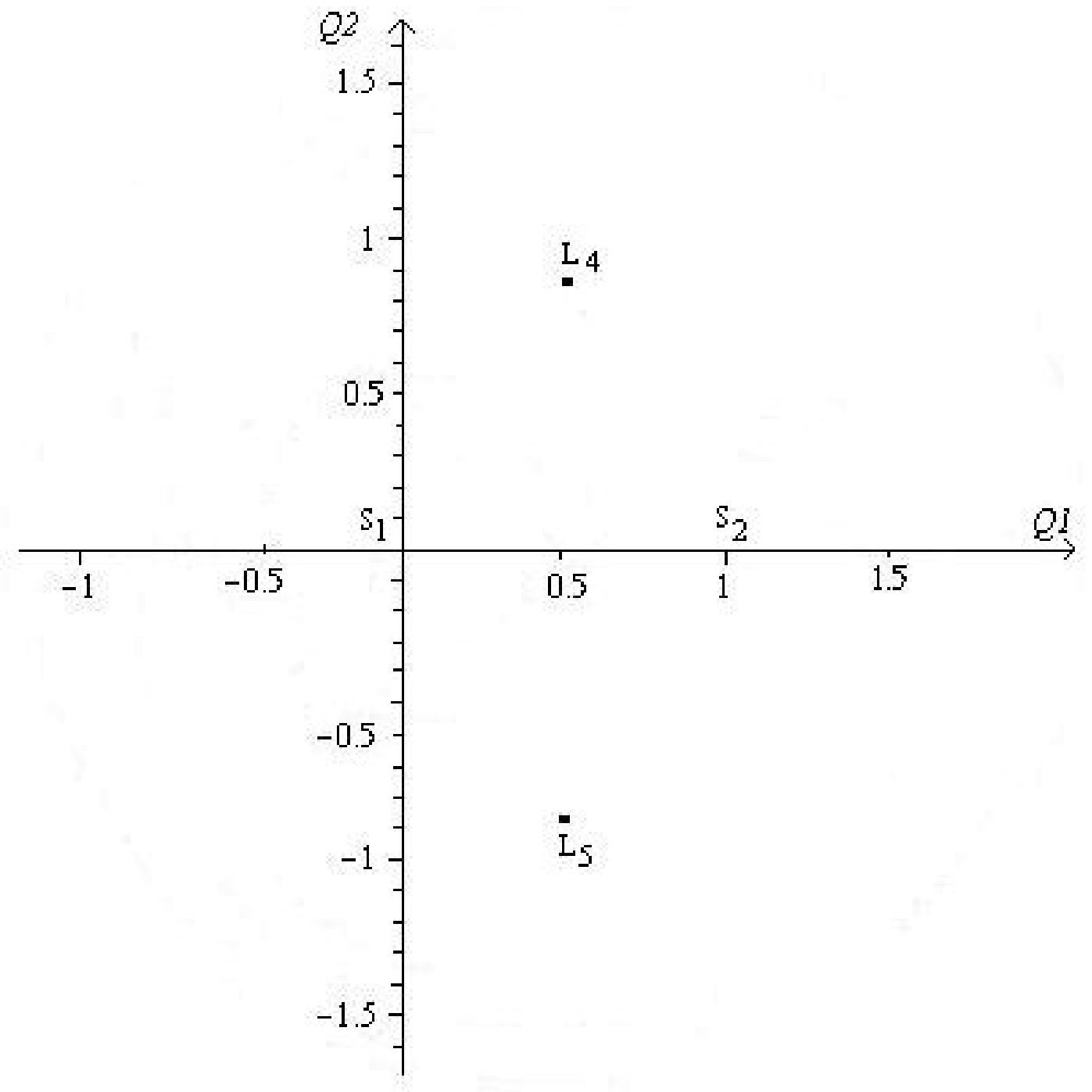}
  \caption{Roche varieties for $n=1$ (in the physical plane)}
  \end{center}
\end{figure}

\begin{figure}
\begin{center}
  \includegraphics[height=0.3\textheight]{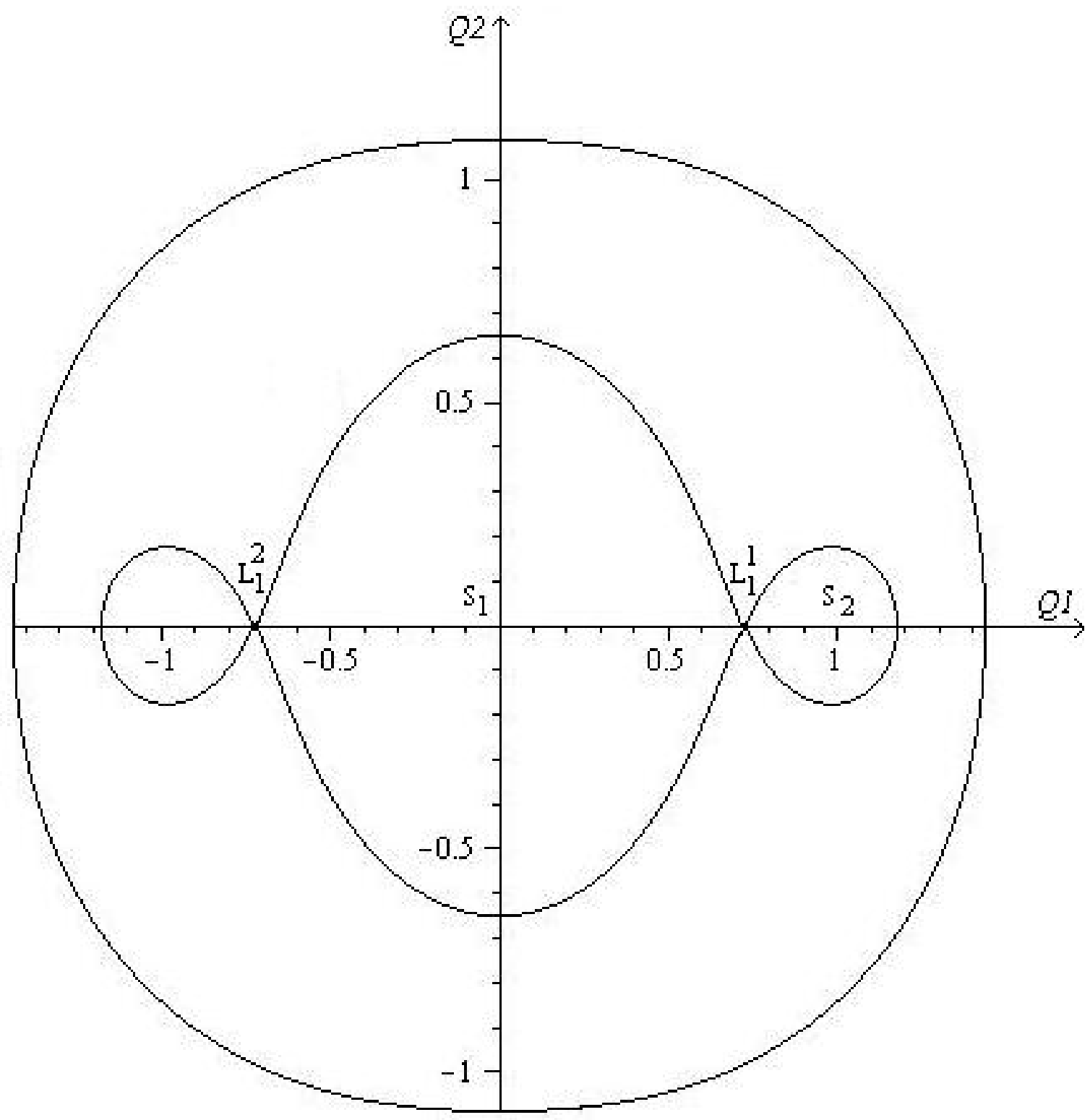}
    \includegraphics[height=0.3\textheight]{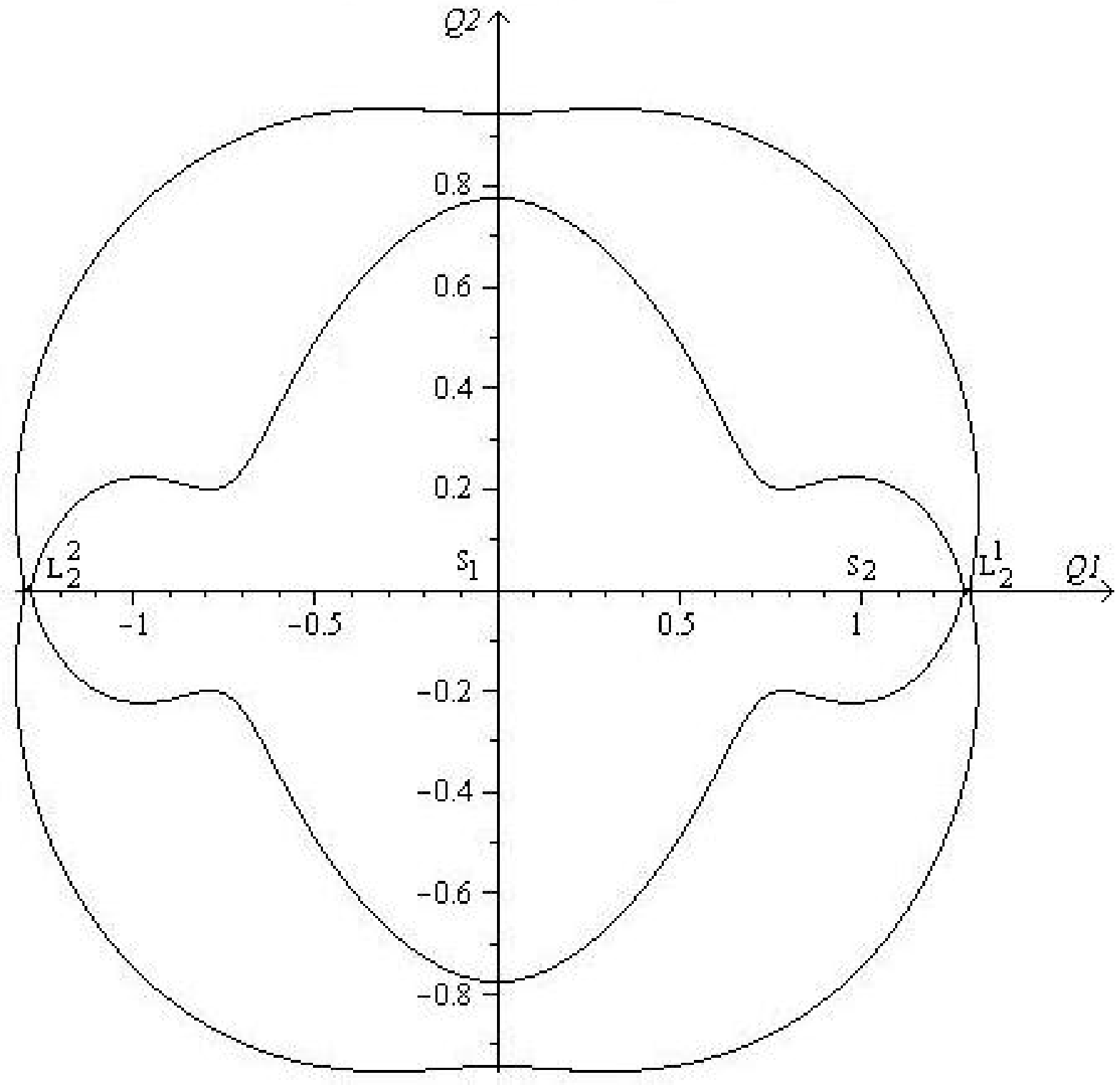}
    \includegraphics[height=0.3\textheight]{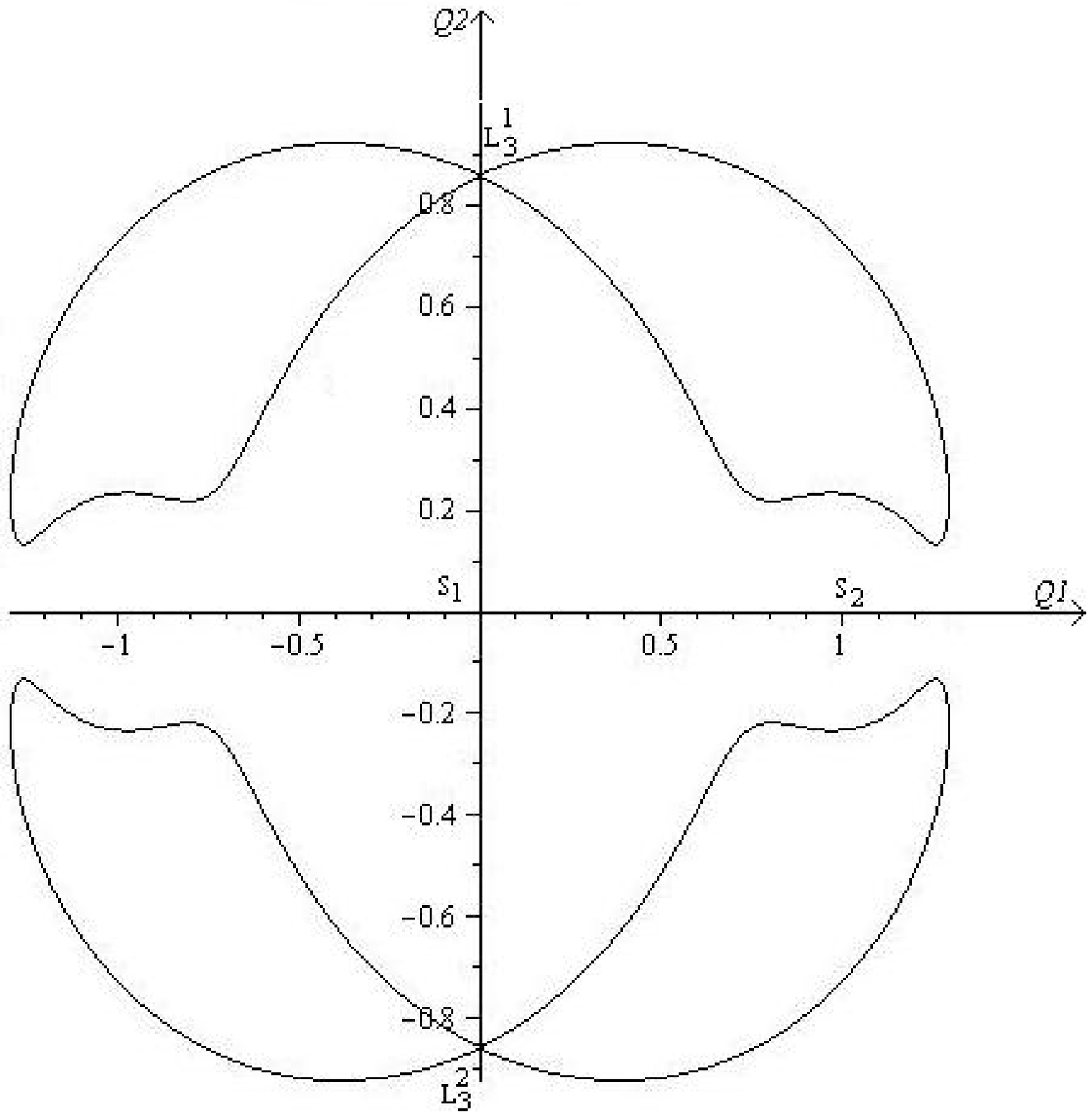}
    \includegraphics[height=0.3\textheight]{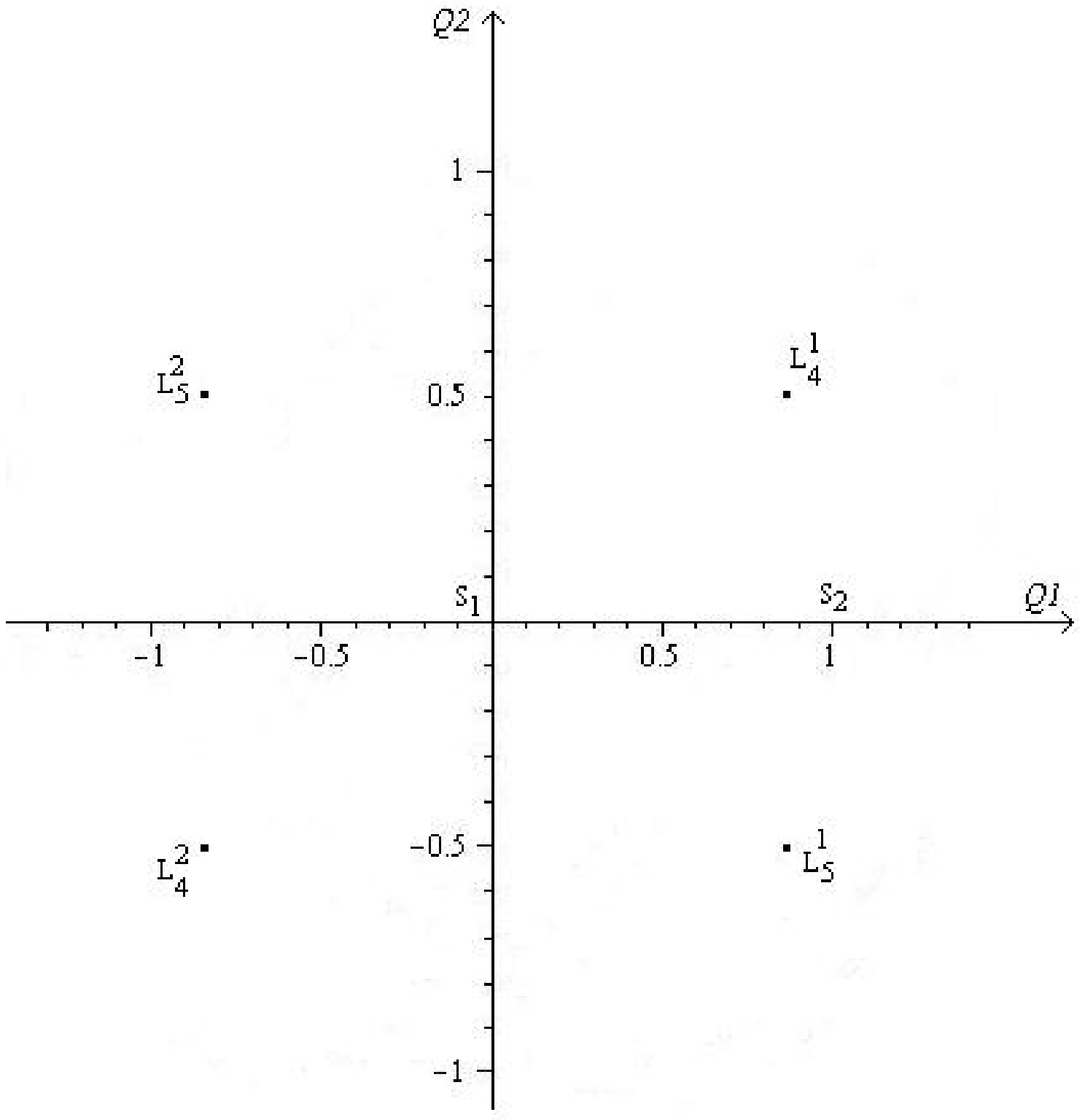}
  \caption{Roche varieties for $n=2$ (Levi-Civita transformation)}
  \end{center}
\end{figure}

\begin{figure}
\begin{center}
 \includegraphics[height=0.3\textheight]{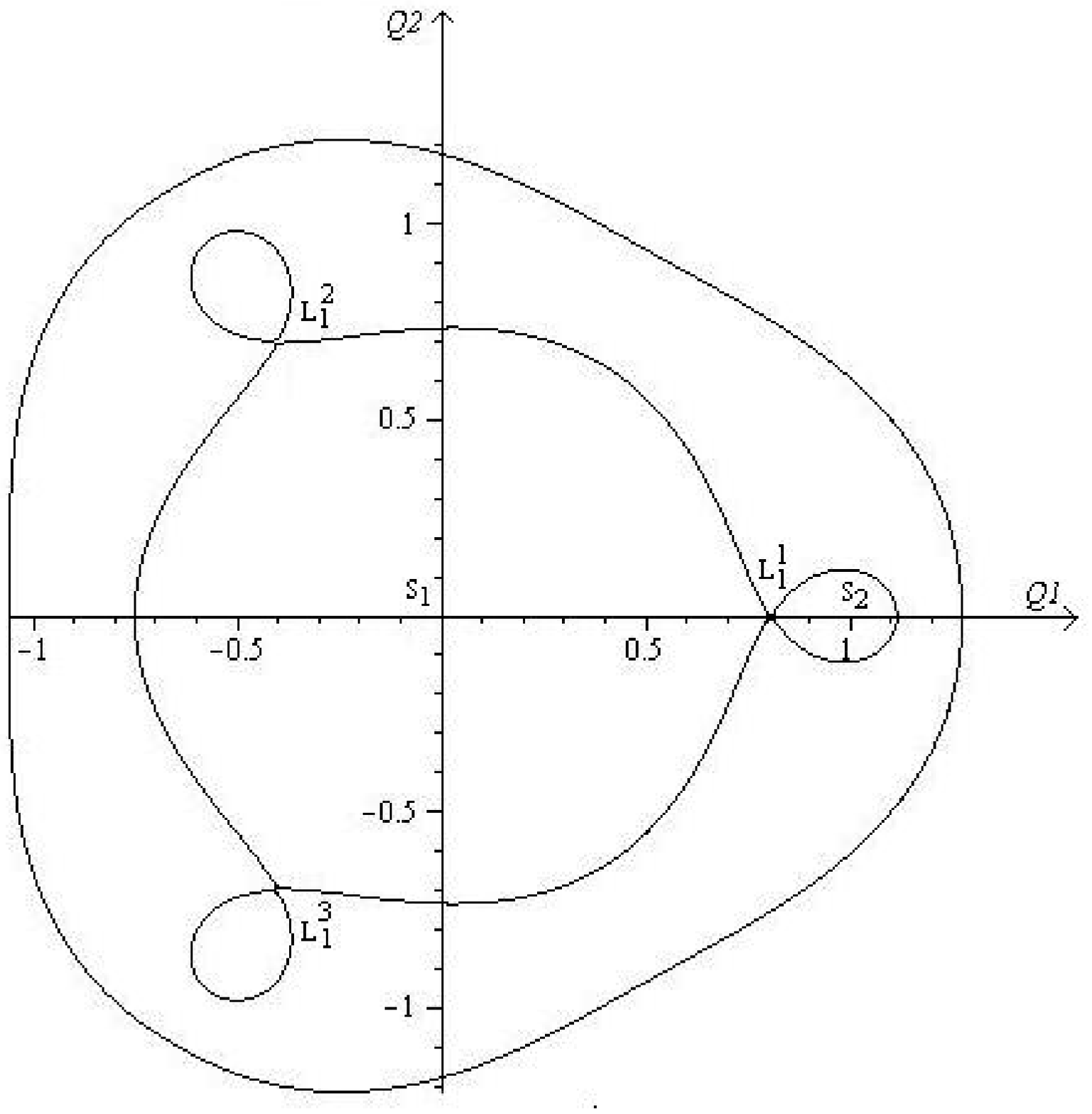}
    \includegraphics[height=0.3\textheight]{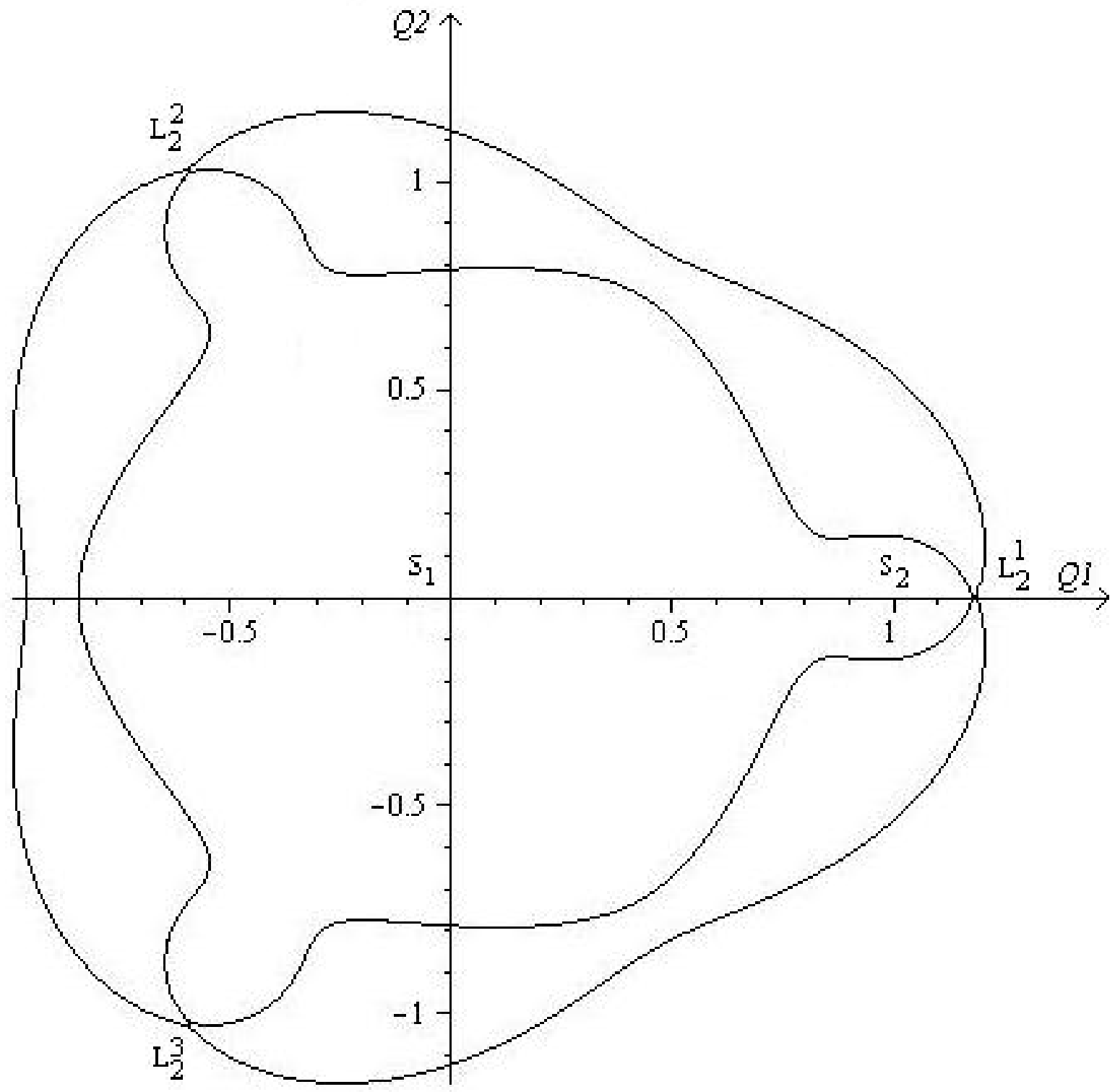}
    \includegraphics[height=0.3\textheight]{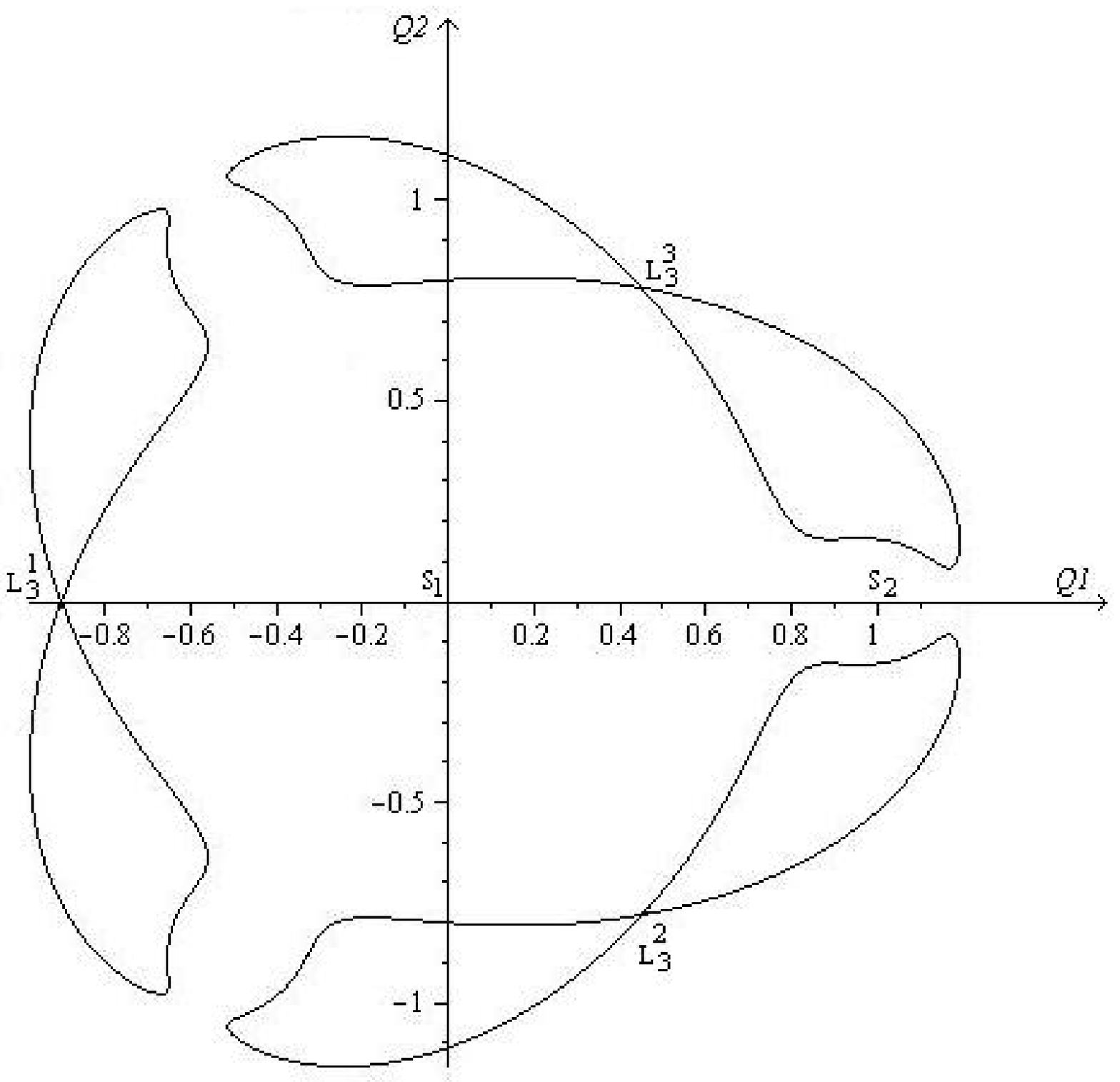}
  \includegraphics[height=0.3\textheight]{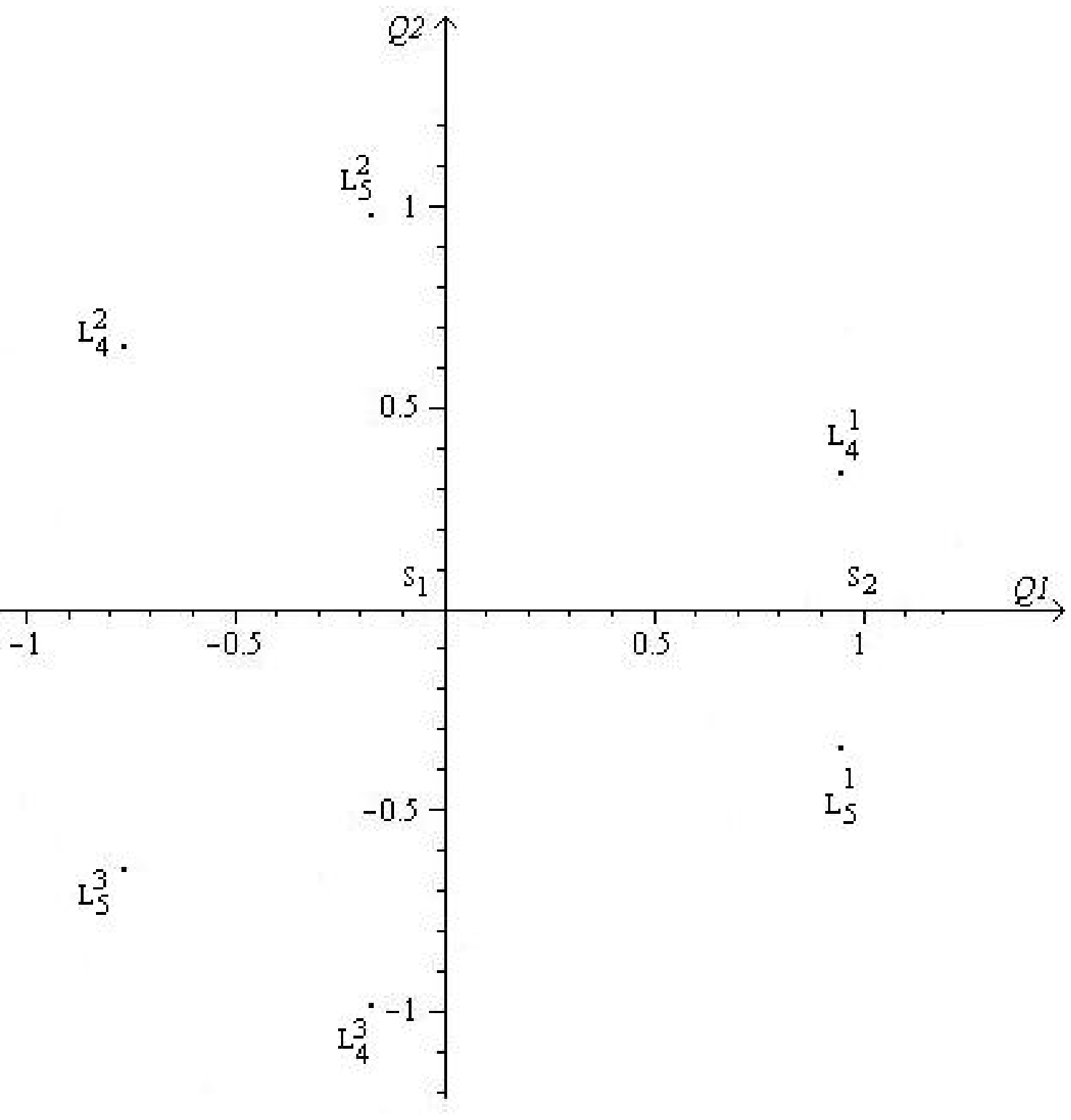}
  \caption{Roche varieties for $n=3$ }
 \end{center}
\end{figure}

\begin{figure}
\begin{center}
 \includegraphics[height=0.3\textheight]{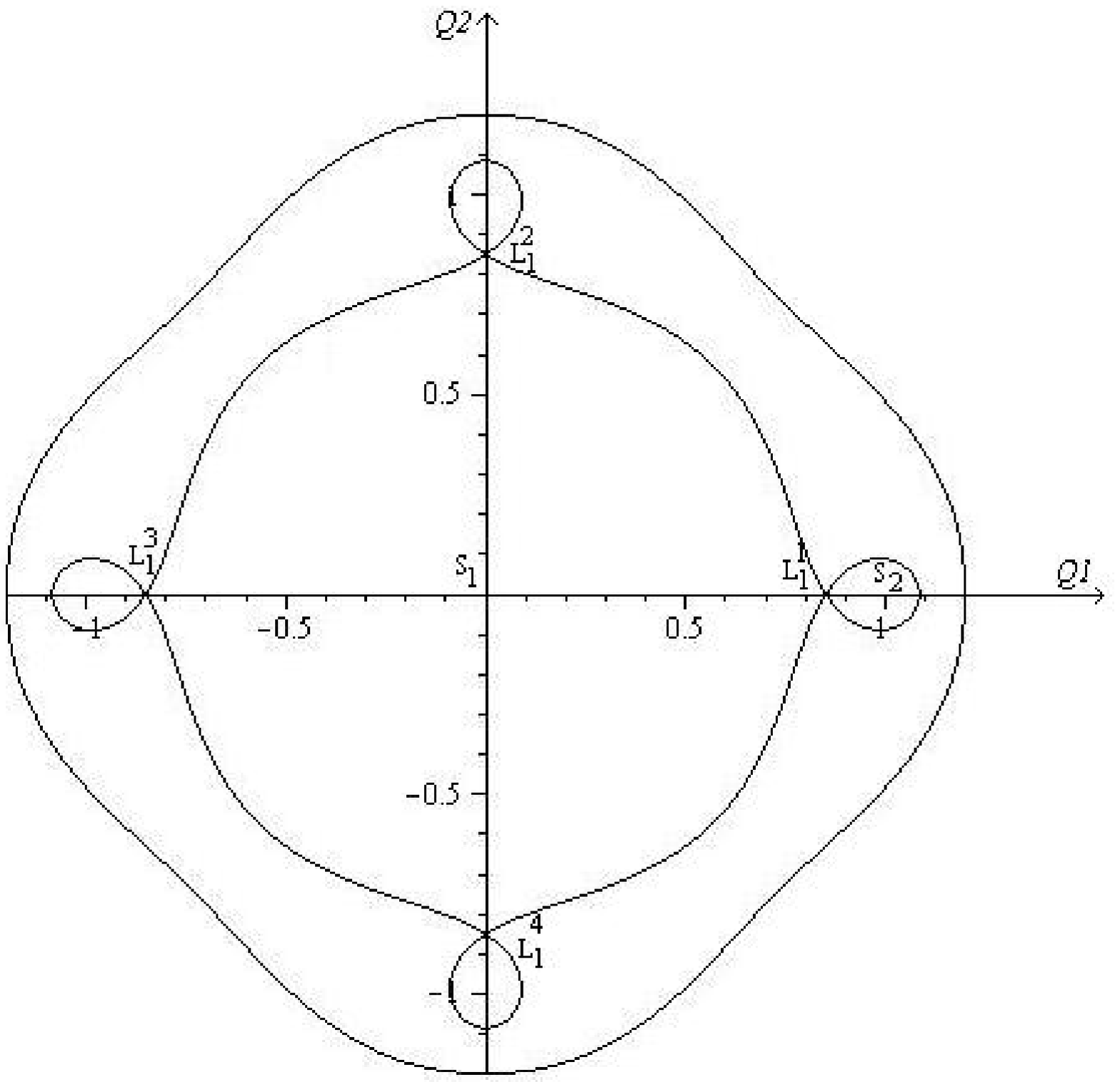}
    \includegraphics[height=0.3\textheight]{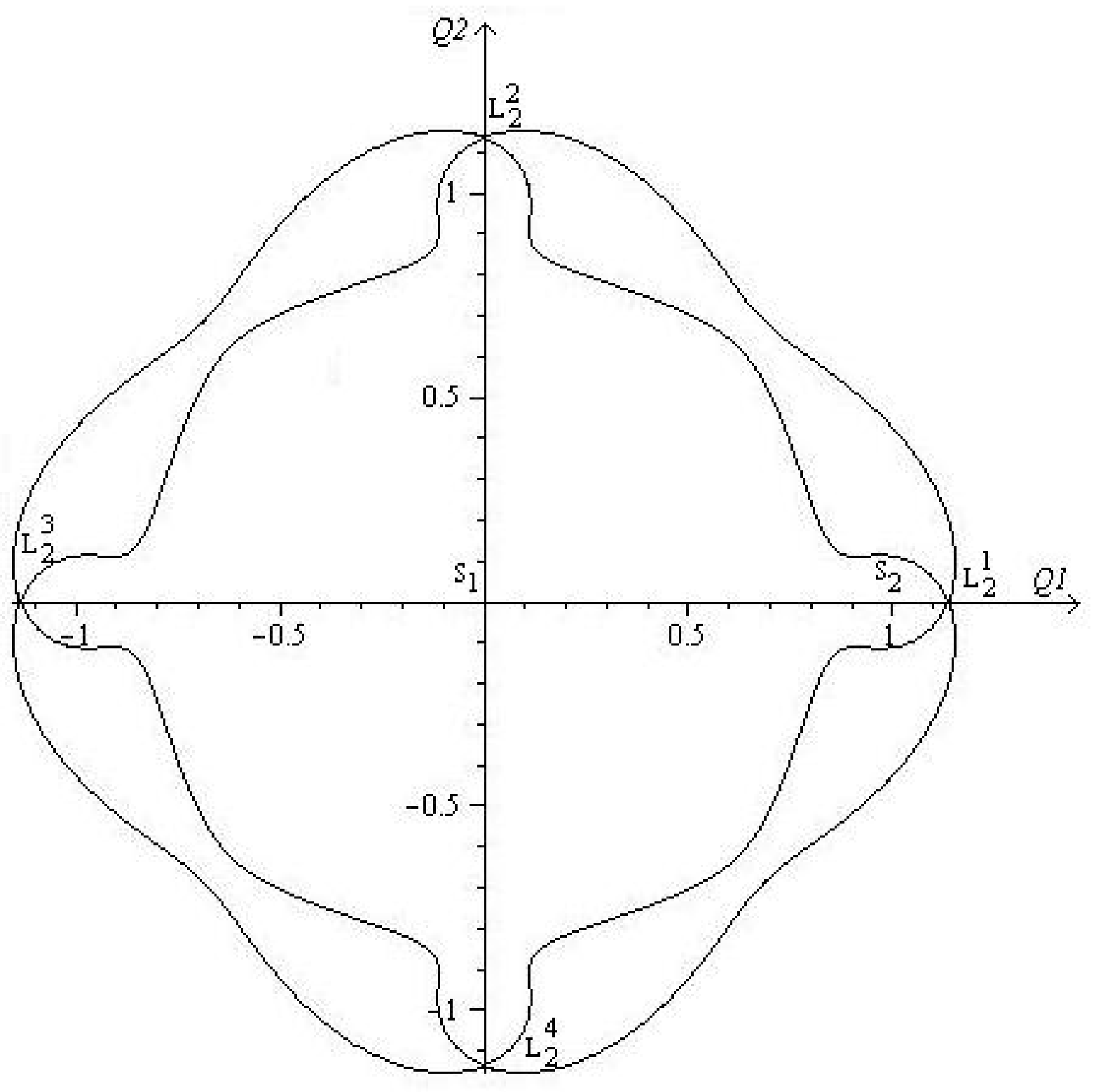}
    \includegraphics[height=0.3\textheight]{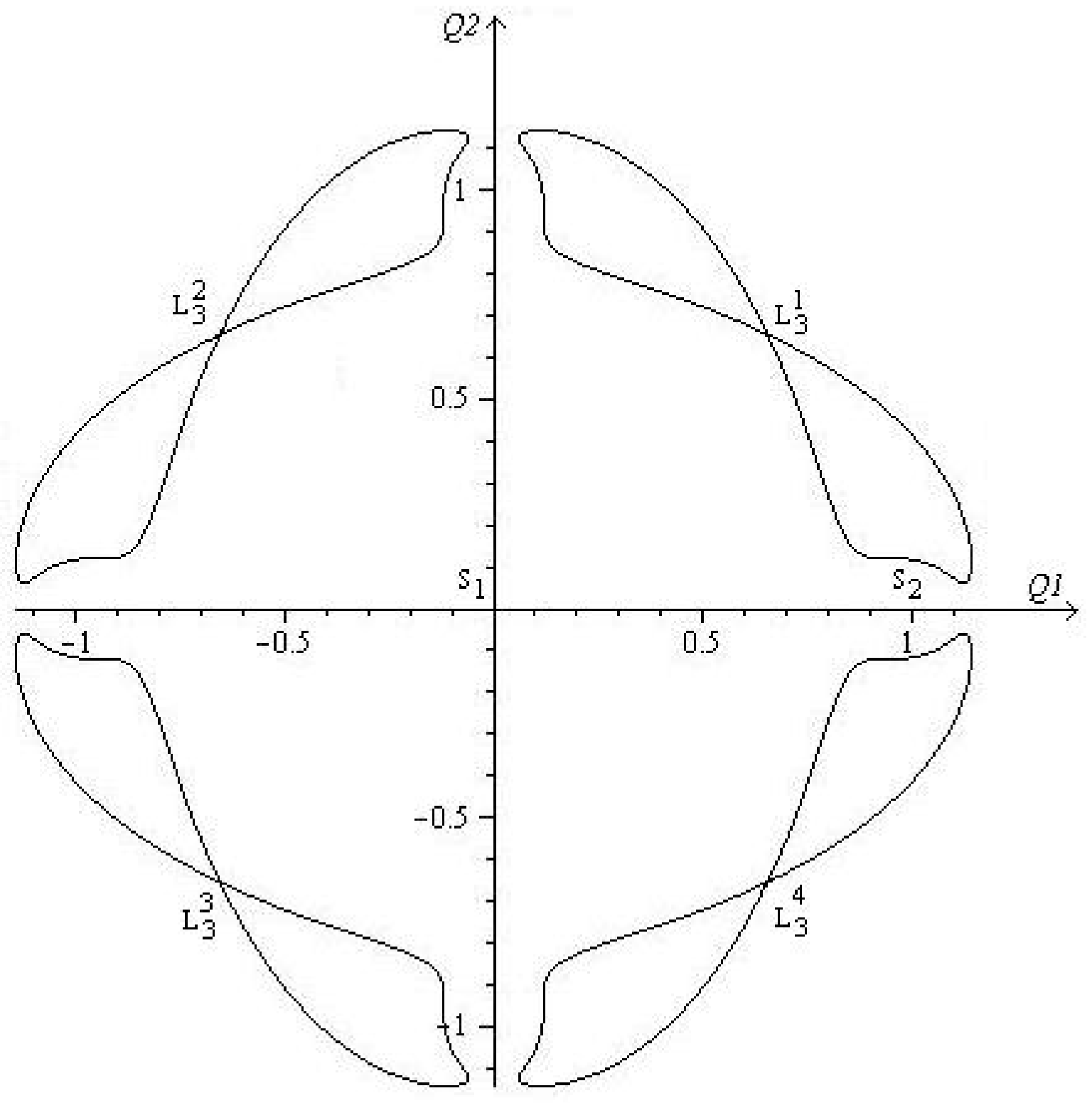}
  \includegraphics[height=0.3\textheight]{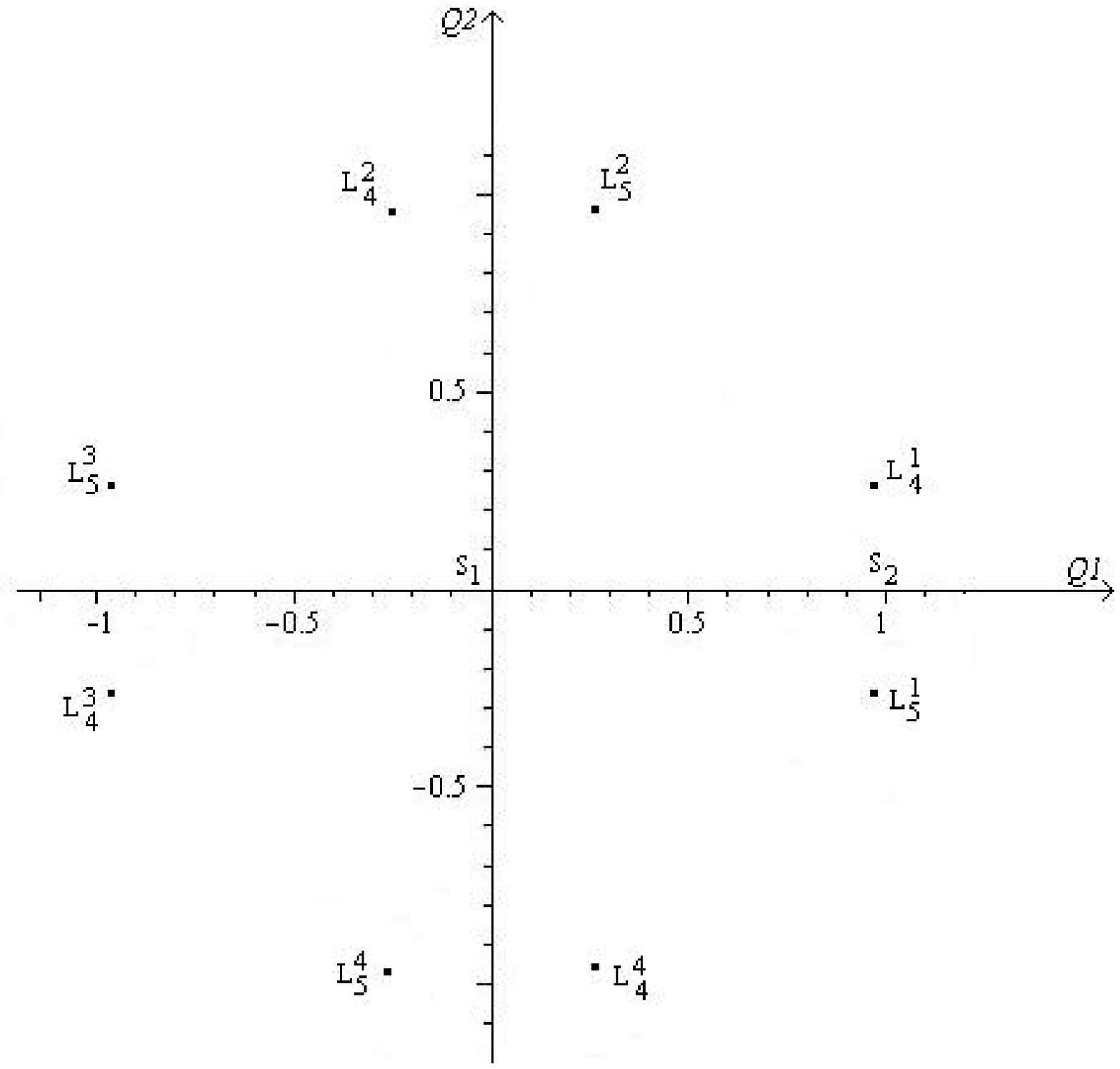}
  \caption{Roche varieties for $n=4$ }
 \end{center}
\end{figure}

\begin{figure}
\begin{center}
 \includegraphics[height=0.3\textheight]{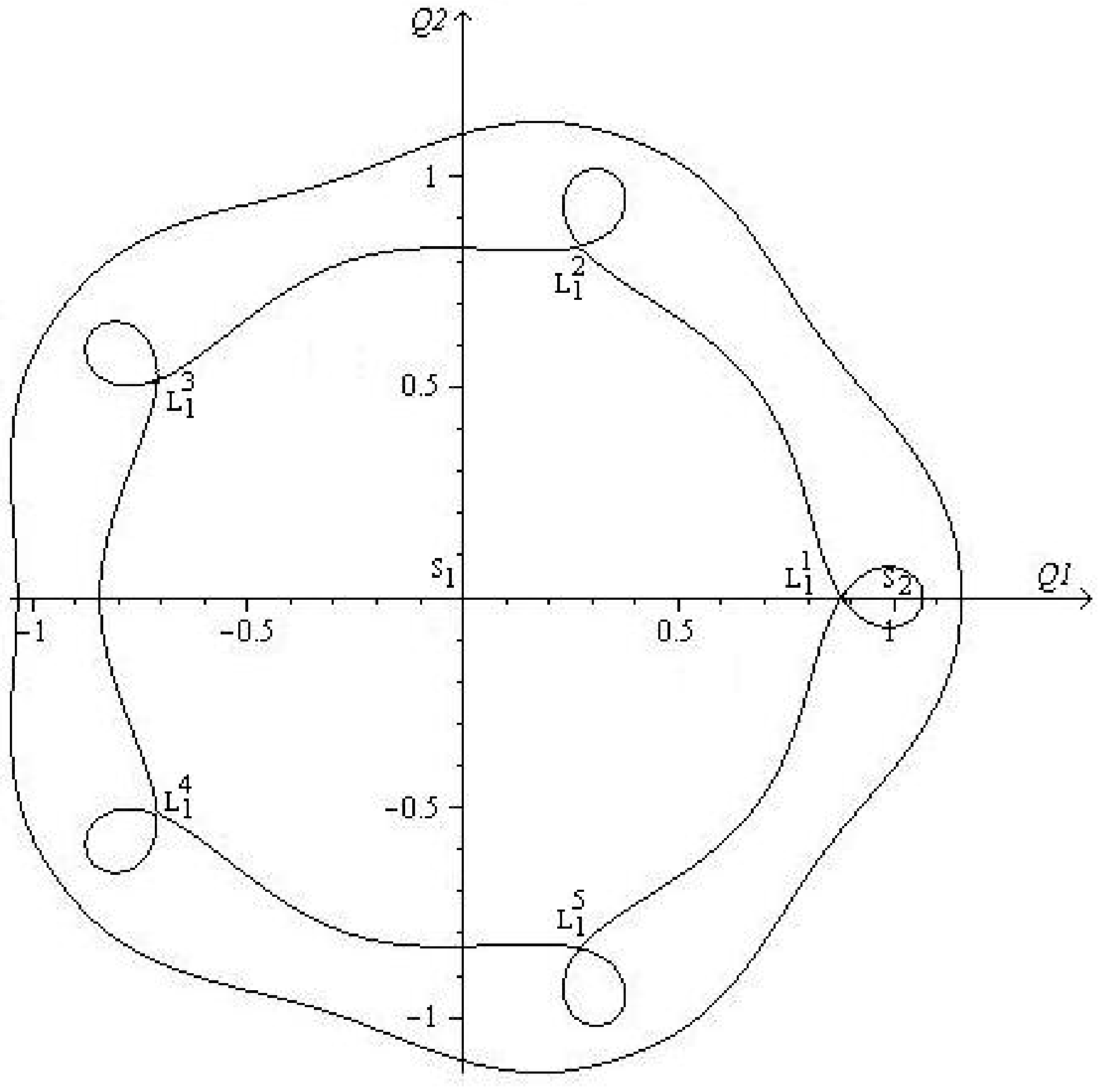}
    \includegraphics[height=0.3\textheight]{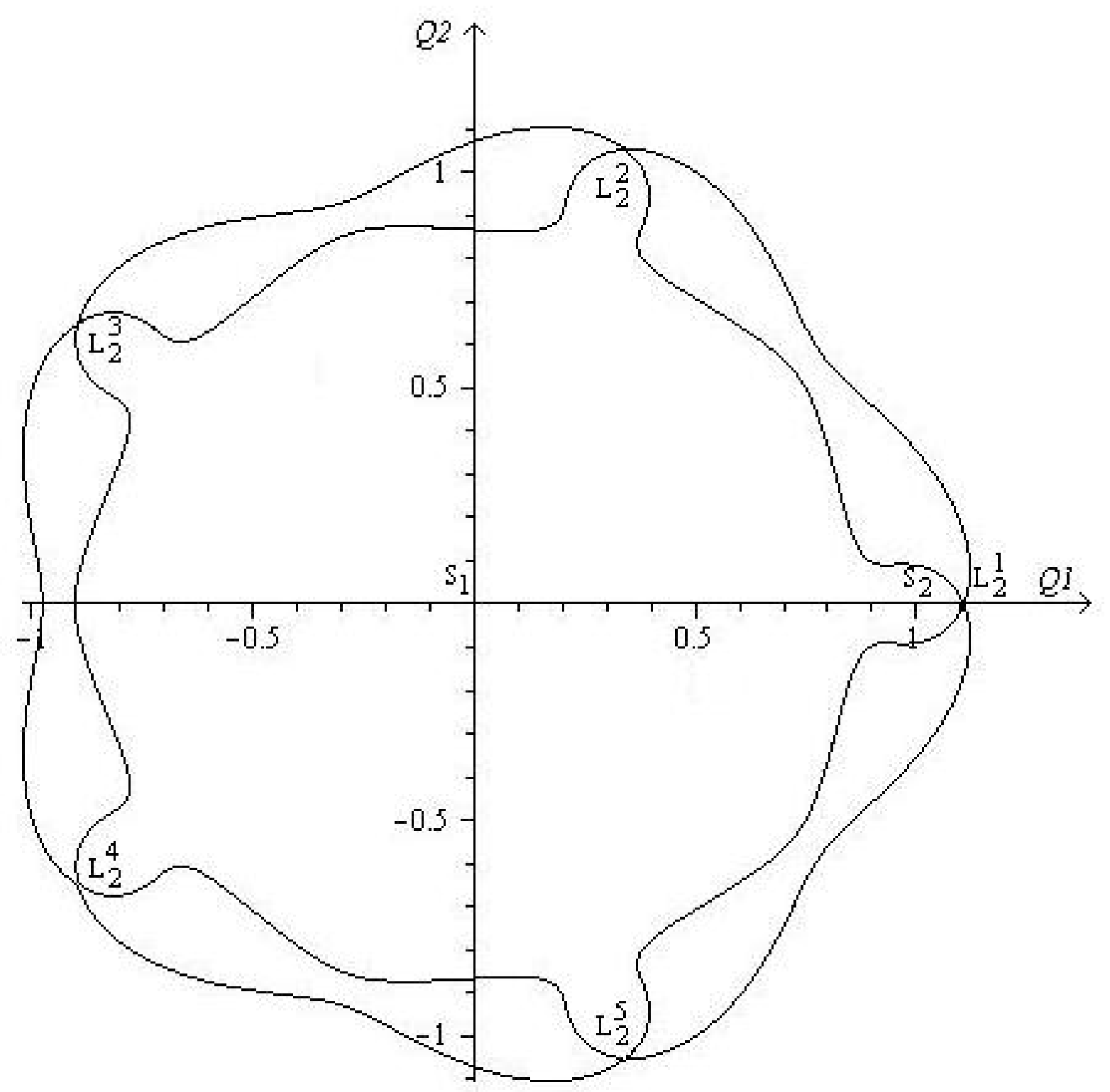}
    \includegraphics[height=0.3\textheight]{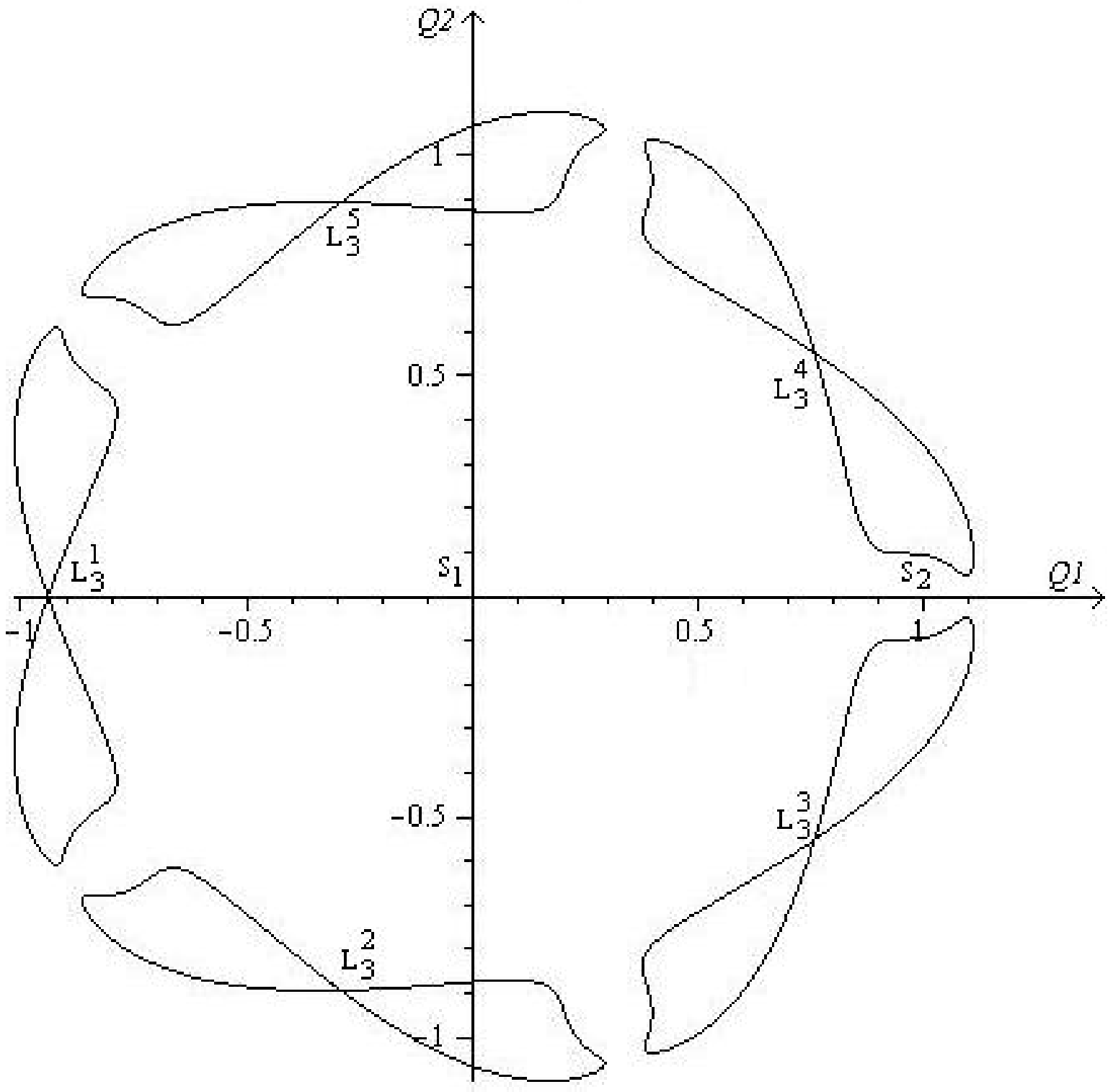}
  \includegraphics[height=0.3\textheight]{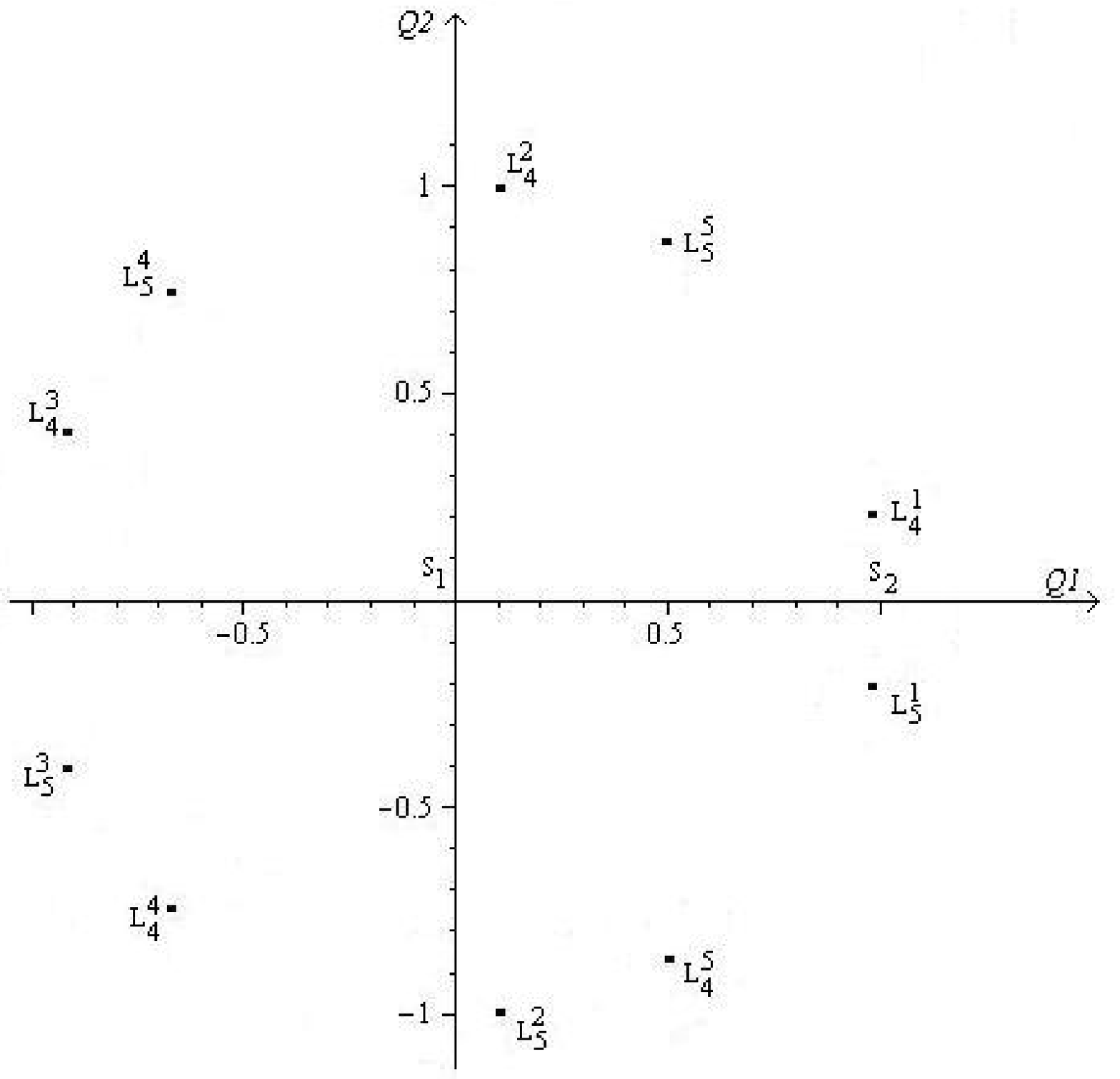}
  \caption{Roche varieties for $n=5$ }
 \end{center}
\end{figure}

\begin{figure}
\begin{center}
 \includegraphics[height=0.3\textheight]{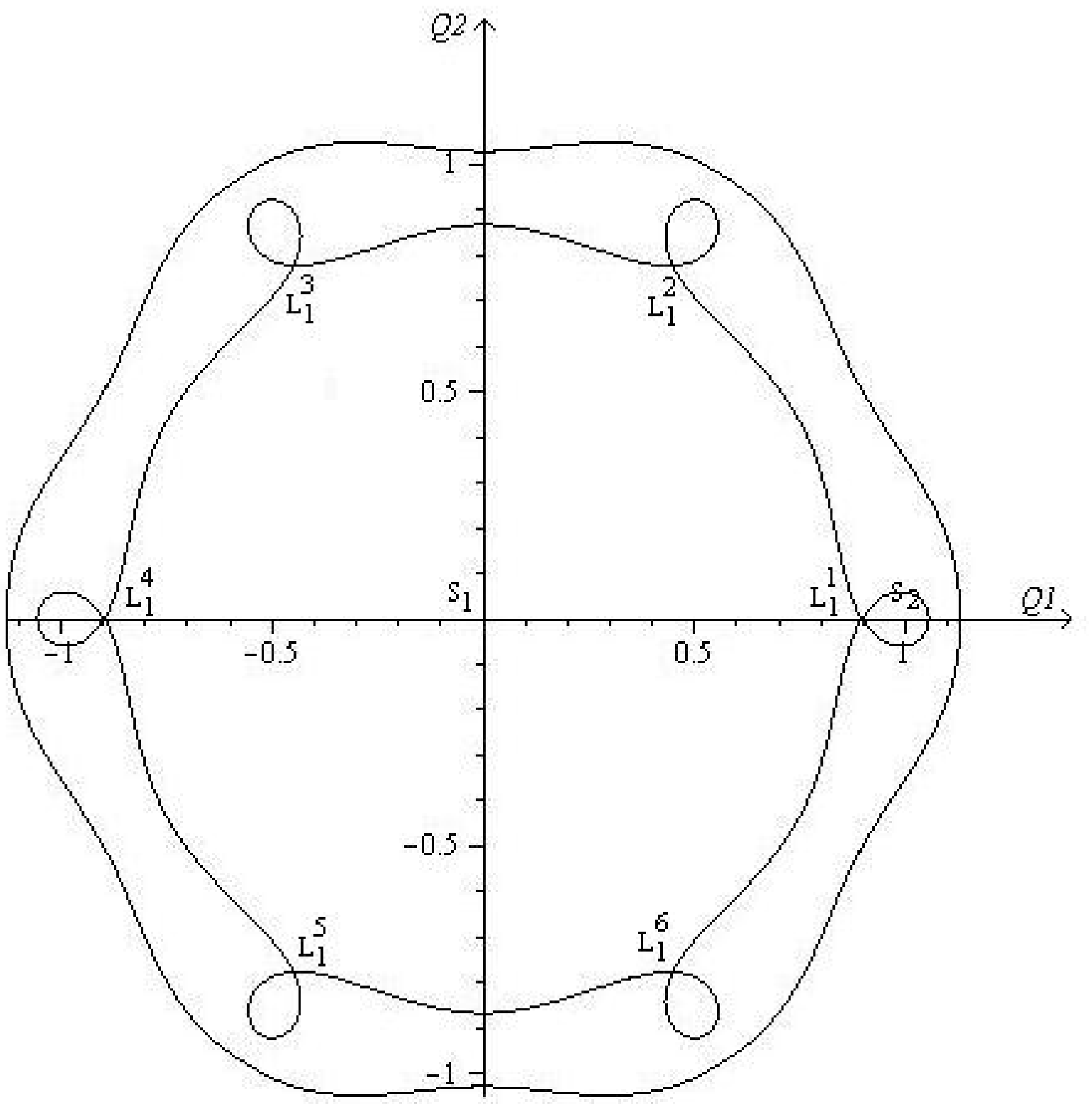}
    \includegraphics[height=0.3\textheight]{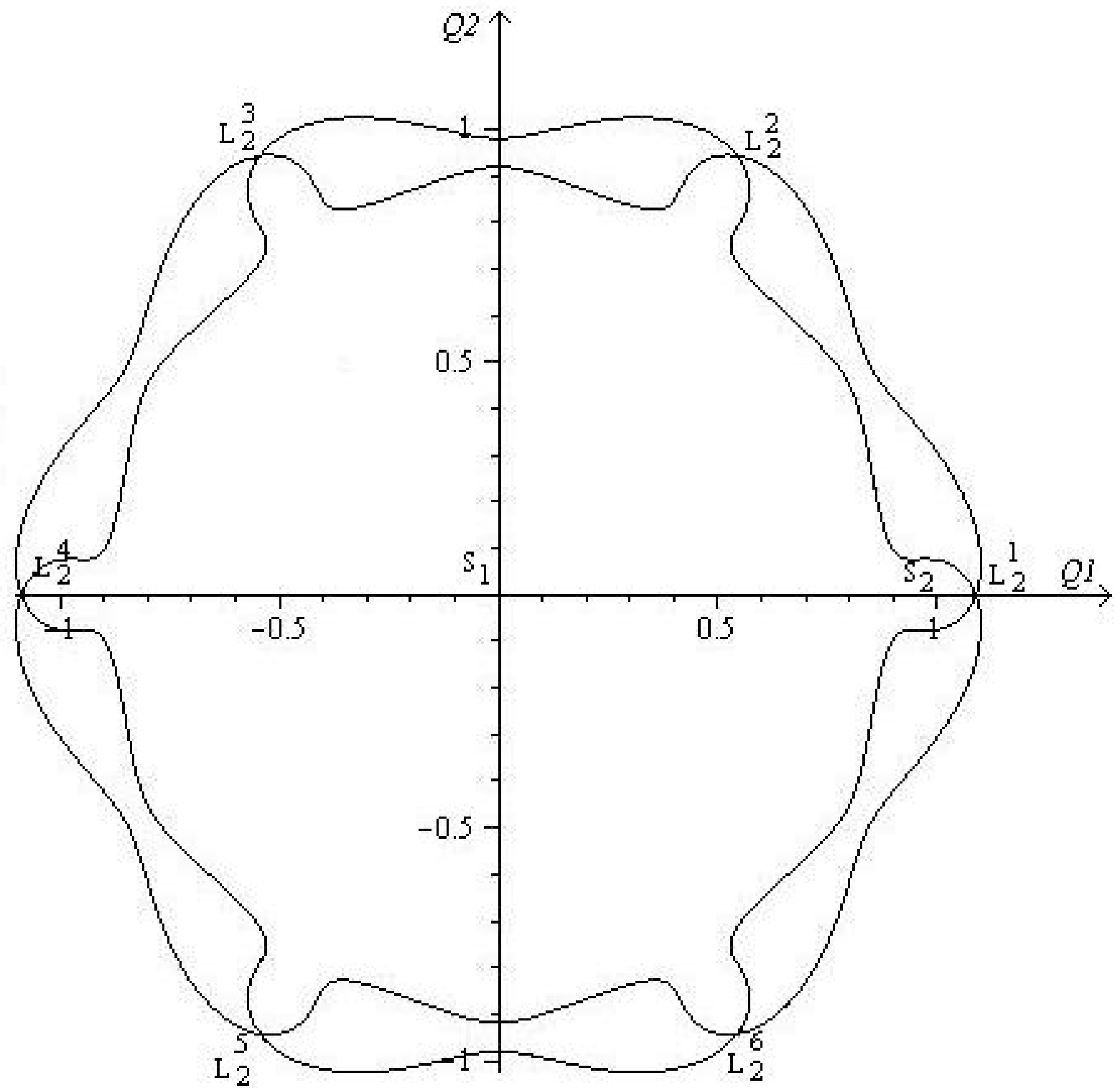}
    \includegraphics[height=0.3\textheight]{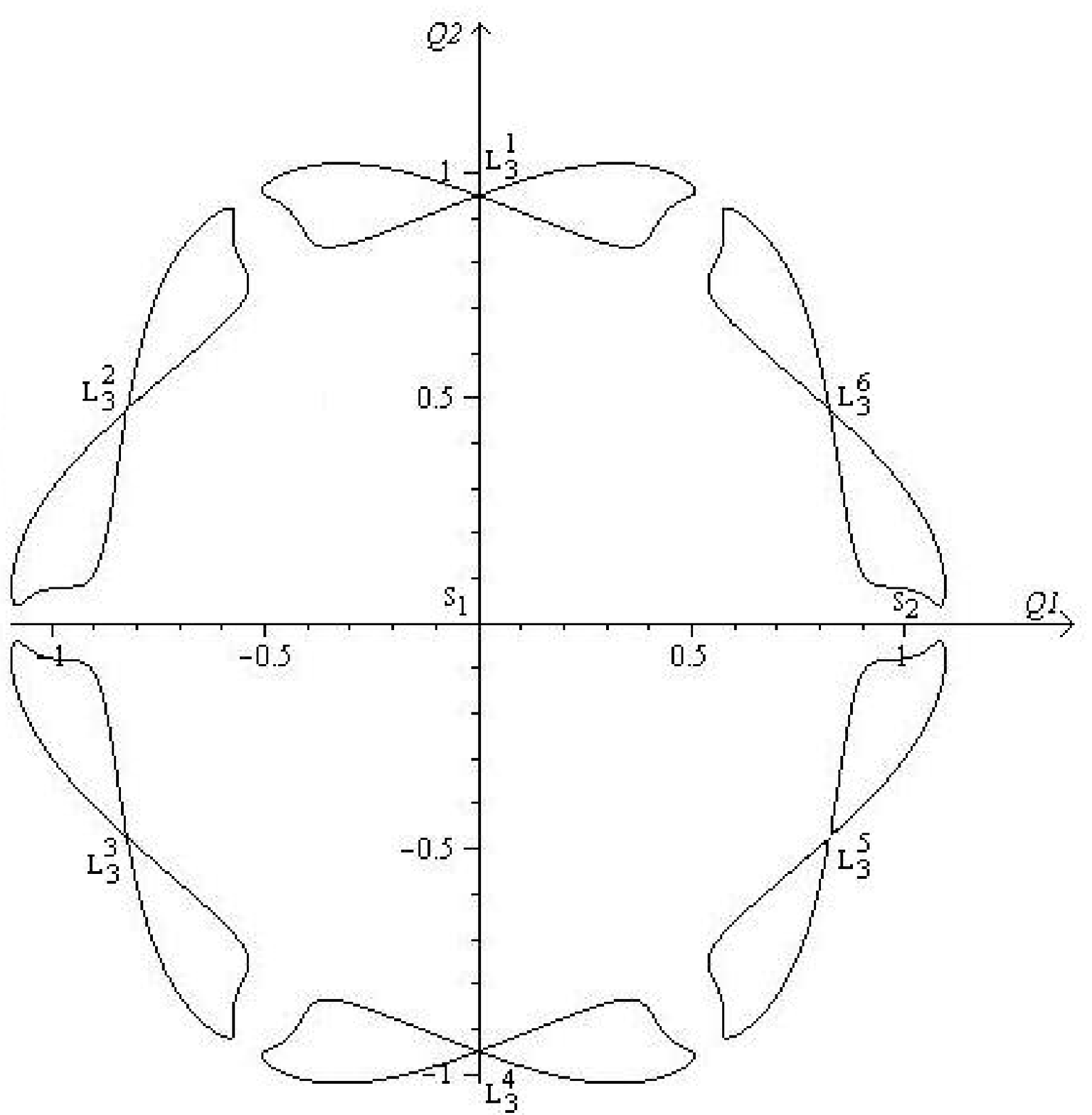}
  \includegraphics[height=0.3\textheight]{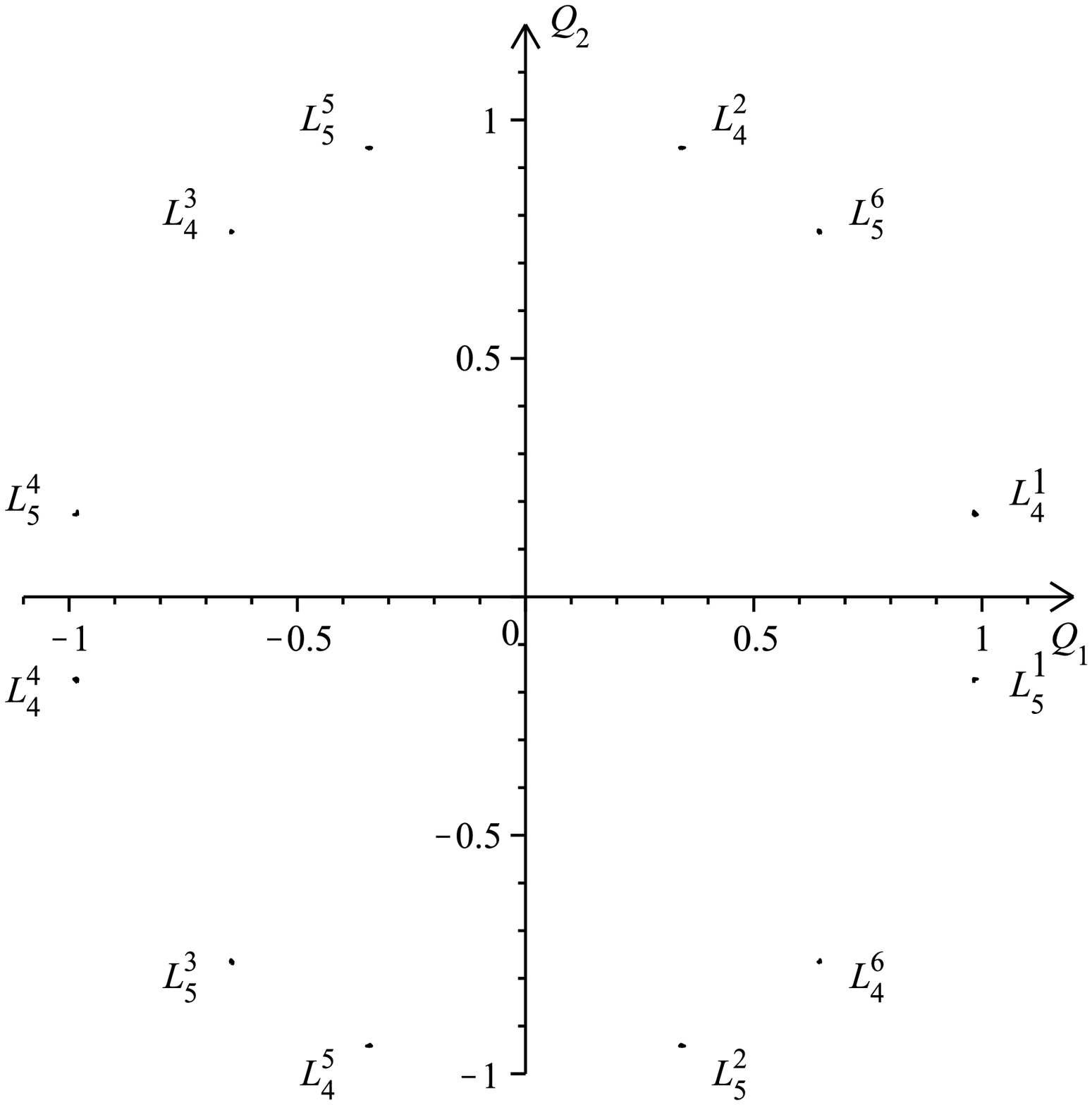}
  \caption{Roche varieties for $n=6$ }
 \end{center}
\end{figure}

\textit{Remarks 2}:
\begin{enumerate}
    \item  For $n=1$ we obtained the equipotential Roche curves in the physical plane.
    \item  For $n=2$ we obtained the Roche varieties in the parametric plane, for the coordinate transformation of Levi-Civita. It is interesting to compare them with those obtained by Szebehely \citep{Sze1966,Sze1967}. At Szebehely the points $L_1^1\;,\;L_1^2\;,\;L_2^1\;,\;L_2^2$ are situated on the ordinate axis, and $L_3^1\;,\;L_3^2$ on the abscissa axis. In our article is reversed. This situation is normal and is due to the different location of the origin of the coordinate system: at Szebehely the origin is located in the mass center of the binary system, and in our article is located in the center of the most massive star $S_1$. In fact, many authors use the barycentric coordinate system, there are important books and articles in which the authors use the coordinate system with origin in the center of the most massive star (\cite{Kopal1978}, \cite{Pla1964}, \cite{Eggleton1983}, \cite{Morris1994}, \cite{Seidov2004}, \cite{Mochnacki1984}).
In Figures 2-6 one can see that the change of the location of the origin of coordinate system has only the effect of rotation of our family of zero-velocity curves, the shape of curves being determined only by the mass ratio of the binary system. We denote this mass ratio with $q$ as in articles cited above. Other authors use $\frac{m_2}{m_1+m_2}$ as parameter (see \cite{Sze1967}), but of course the shape is the same, having in view that $\mu=\frac{q}{1+q}$.

        \item The Jacobi constant signifies energy, so is normal to be the same in the physical and parametric plane. We can demonstrate this analytically. We denote in the following with $C_{L_1}$ the Jacobi constant in the physical plane corresponding to the equilibrium point $L_1$, and $C_{L_1^1}$  the Jacobi constant in the parametric plane corresponding to the equilibrium point $L_1^1$, and we will demonstrate that  $C_{L_1}=C_{L_1^1}$.
        The equilibrium points in the physical plane can be obtained from the system (18)-(19). For $L_1({q_1}_{L_1}, \;0)$ we have  ${q_2}_{L_1}=0$, $r_1={q_1}_{L_1},\;\;r_2=1-{q_1}_{L_1}$. Then, the equation which gives the abscissa of $L_1$ is:
        \begin{equation}
        \left({q_1}_{L_1}-{q_1}_{L_1}^{-2}\right)-q\left[ (1-{q_1}_{L_1})-(1-{q_1}_{L_1})^{-2} \right]=0.
       \end{equation}
    The equilibrium points in the parametric plane can be obtained from the system (22)-(23).  For $L_1^1({Q_1}_{L_1^1}, \;0)$ we have  ${Q_2}_{L_1^1}=0$, $r_1=({Q_1}_{L_1^1})^n,\;\;r_2=1-({Q_1}_{L_1^1})^n$. Then, the equation which gives the abscissa of $L_1^1$ is:
\begin{equation}
        \left[({Q_1}_{L_1^1})^n-(({Q_1}_{L_1^1})^n)^{-2}\right]-q\left[ (1-({Q_1}_{L_1^1})^n)-(1-({Q_1}_{L_1^1})^n)^{-2} \right]=0.
    \end{equation}
        Comparing equations (24) and (25), we observe that ${q_1}_{L_1}=({Q_1}_{L_1^1})^n=a$. From eq. (9) we have for the Jacobi constant in the physical plane:
        $$C_{L_1}=a^2-\frac{2qa}{1+q}+\frac{2}{(1+q)a}+\frac{2q}{(1+q)(1-a)},$$
and from eq. (20) we have for the Jacobi constant in the parametric plane:
        $$C_{L_1^1}=a^2-\frac{2qa}{1+q}+\frac{2}{(1+q)a}+\frac{2q}{(1+q)(1-a)}.$$
So, $ C_{L_1}=C_{L_1^1}.$
Similar for $C_{L_i}=C_{L_i^1},\;\;i\in \{ 2,3,4,5 \}.$
\end{enumerate}

\section{Asymptotic variety in the parametric plane}

Let us write the eq. (9) in the form:
\begin{equation}
q_1^2-2\frac{qq_1}{1+q}+\left(\frac{q}{1+q}  \right)^2+q_2^2+\frac{2}{(1+q)r_1}  +\frac{2q}{(1+q)r_2} \;=\;C+\left(\frac{q}{1+q} \right)^2,
\end{equation}
where $r_1=\sqrt{q_1^2+q_2^2},\;\;\;\;r_2=\sqrt{(q_1-1)^2+q_2^2}.$

For big values of $q_1$ and $q_2$ satisfying this equation, the fifth and sixth terms in the left side are relatively unimportant, and the equation can be write:
$$ \left(q_1-\frac{q}{1+q}\right)^2+q_2^2=C+\left( \frac{q}{1+q} \right)^2-\frac{2}{(1+q)r_1}-\frac{2q}{(1+q)r_2}=C+\left(  \frac{q}{1+q} \right)^2-\varepsilon, $$
where $\varepsilon$ is a small quantity. This is the equation of the asymptotic circle in the physical plane; it has the center located in the point $D(\frac{q}{1+q},0)$ and the radius $\sqrt{C+\left(  \frac{q}{1+q} \right)^2-\varepsilon}$. It can be compared with the equation of the asymptotic circle from Moulton (see \cite{Mou1923}), where the center of the coordinate system is located in the mass center of the binary system.

Considering now in the parametric plane the eq. (12), we obtain for big values of $Q_1$ and $Q_2$, small values of terms $\frac{2}{(1+q)r_1}$ and $\frac{2q}{(1+q)r_2}$, relatively unimportant, where
$r_1=\sqrt{\left( Q_1^2+Q_2^2 \right)^n} $, and $r_2=\sqrt{\left( Q_1^2+Q_2^2 \right)^n-2\Re\left[ (Q_1+iQ_2)^n \right]+1}$. Then the eq. (12) can be written:
\begin{equation}
(Q_1^2+Q_2^2)^n-2\frac{q\Re{(Q_1+iQ_2)^n}}{1+q}=C-\varepsilon,
\end{equation}
with $\varepsilon$ being a small quantity.

Let us denote this curve \textit{the asymptotic variety}. So, by using the generalized Levi-Civita transformation, the asymptotic circle from the physical plane is transformed into the asymptotic variety given by eq. (27).

\textit{Remarks 3}:
\begin{enumerate}
    \item  For $n=1$ we obtained the asymptotic circle, in the physical plane:
                $$\left( Q_1-\frac{q}{1+q}  \right)^2+Q_2^2=C+\left( \frac{q}{1+q}  \right)^2-\varepsilon.$$
    \item  For $n>1$ the asymptotic variety is no longer a circle.
\end{enumerate}

In the following we will analyze this situation. In a previous article we demonstrated the following theorem \citep{RR2014}:

\textbf{Theorem 2.}

\textit{In the polynomial regularization's methods, if A is an arbitrary point of the trajectory in the physical plane {\rm ($n=1$)}, and B is its corresponding point in the parametric plane, then we have the following relations concerning the polar radii and angles}:
$$|S_1A|=|S_1B|^n,\;\;\;\;  \widehat{AS_1q_1}=n\cdot \widehat{BS_1Q_1},\;\;\;\; \forall\;n\in \mathbb{N},\;\;n\geq1.$$

In Figure 7 there is represented in the left side the asymptotic circle in the physical plane, for $q=0.8$. We took this big value because the corresponding asymptotic circle in the physical plane has the center far from the origin of the coordinate system and the reasoning is easier to follow. The smallest polar radius is $S_1A$, and it corresponds to the polar angles $\widehat{AS_1q_1}=180^{\circ},\;\;1\cdot 360^{\circ}+180^{\circ},\;\;2\cdot 360^{\circ}+180^{\circ},...$.

Let us select a value for $n$, for example $n=3$. Then, in the parametric plane, the smallest polar radius will be $S_1B=\sqrt[3]{S_1A}$, and it will correspond to the polar angles: $\widehat{B^1S_1Q_1}=\frac{180^{\circ}}{3};\;\;\widehat{B^2S_1Q_1}=\frac{1\cdot 360^{\circ}+180^{\circ}}{3};\;\; \widehat{B^3S_1Q_1}=\frac{2\cdot 360^{\circ}+180^{\circ}}{3}$, namely $\widehat{B^1S_1Q_1}=60^{\circ},\;\;\widehat{B^2S_1Q_1}=180^{\circ},\;\;\widehat{B^3S_1Q_1}=300^{\circ}$ (see Figure 7 in the middle.)

\begin{figure}
\begin{center}
 \includegraphics[height=0.3\textheight]{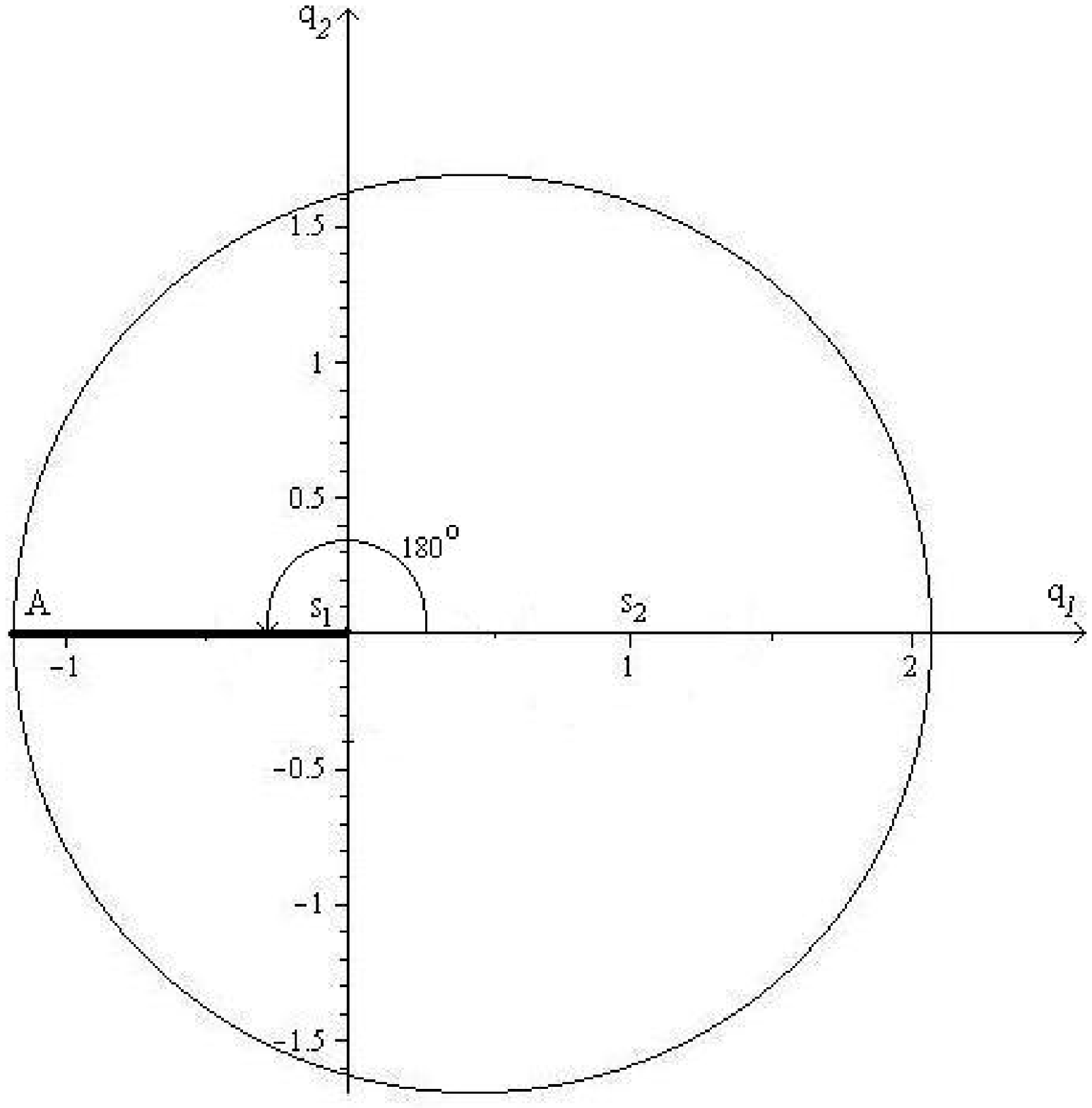}
    \includegraphics[height=0.3\textheight]{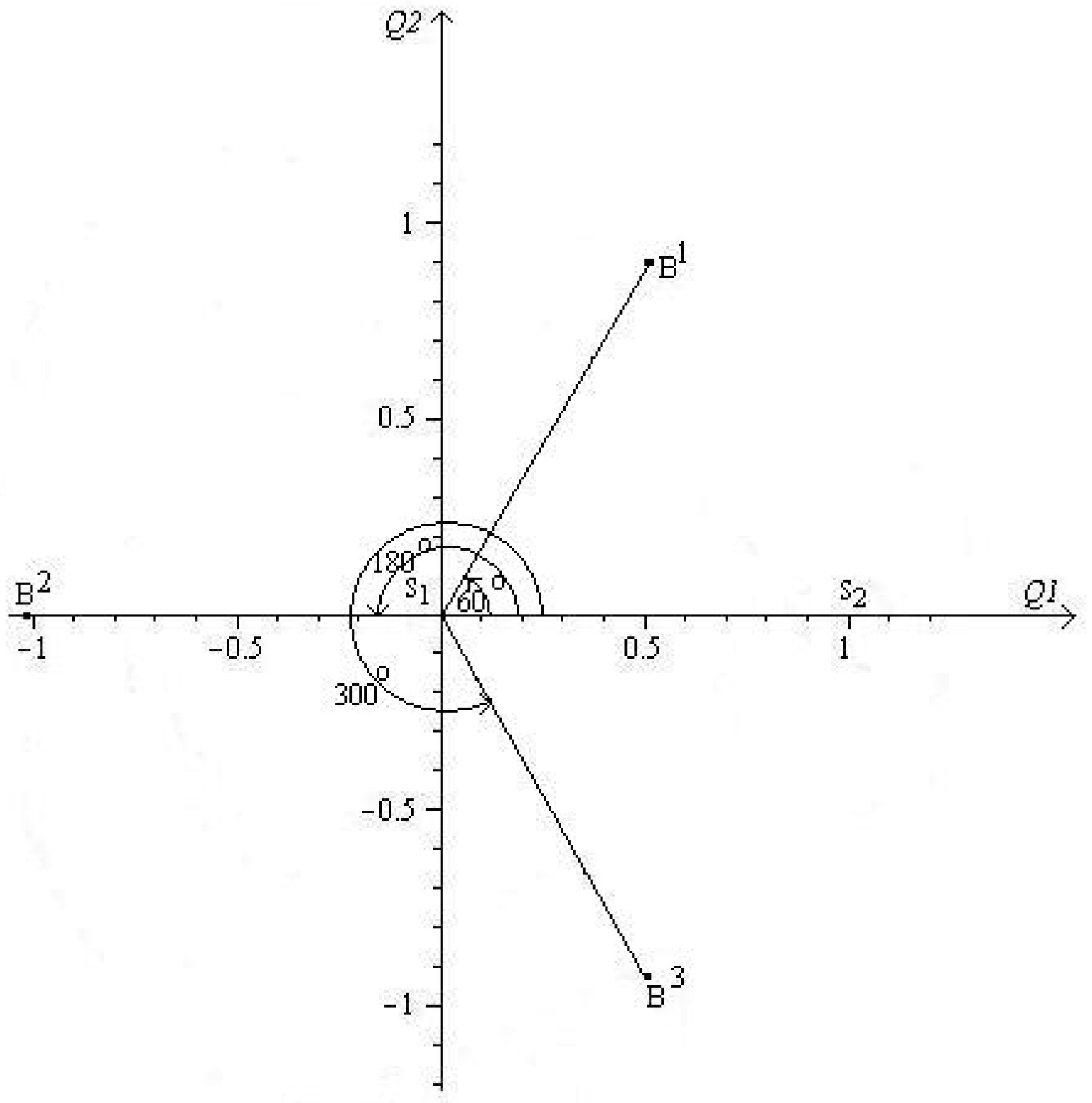}
    \includegraphics[height=0.3\textheight]{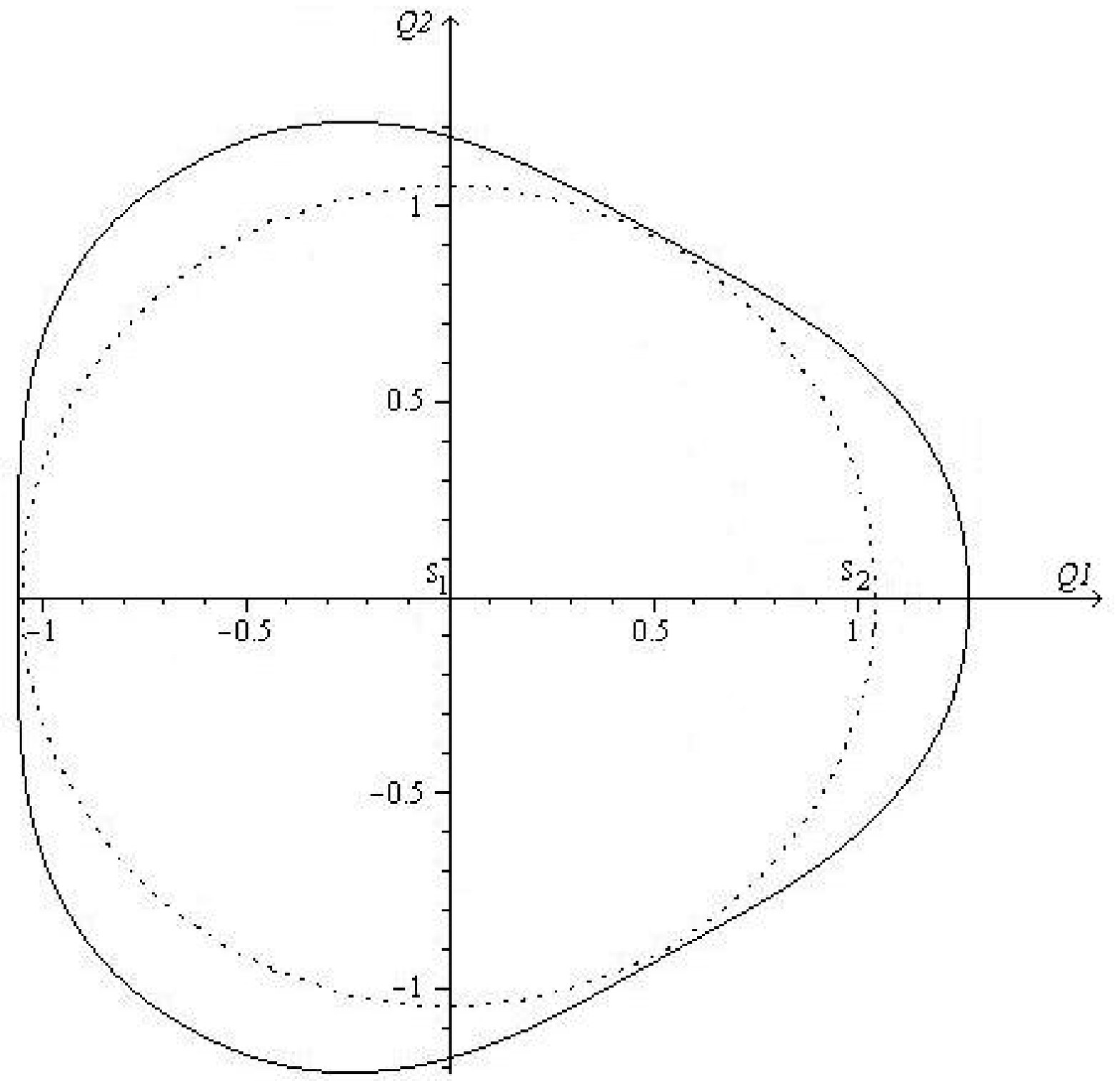}
  \caption{Asymptotic variety}
 \end{center}
\end{figure}

One can remark that, in order to completely cover the parametric plane, it was necessary to browse three times the physical plane. To the point A, which is the nearest point of origin in the physical plane, it correspond three points $B^1,\;\;B^2,\;\;B^3$, the nearest points of origin in the parametric plane. In Figure 7 (below),  is plotted with solid line the asymptotic variety, and with point-line the circle determined by the points $B^1,\;\;B^2,\;\;B^3$; one can see that the circle has a smaller (or equal) radius than the polar radius of the asymptotic variety.

Therefore the asymptotic variety in the parametric plane is not a
circle, but it has $n$ inlets (see Figure 2,...,6 the left hands,
up). This situation is due to the fact that the origin in the
physical plane is taken in the center of the most massive star,
not in the mass center of the binary system.

\section{The slope of Roche variety in $L_{1}^1$ point}

The slope of the curve of zero velocity into the point $L_1$ in the physical plane is given by (see \cite{Pla1964}, \cite{RR2003}, and Figure 1):
\begin{equation}
\tan ^2\phi=-\frac{\frac{\partial ^2 \psi}{\partial q_1^2}}{\frac{\partial ^2 \psi}{\partial q_2^2}},
\end{equation}
where $\psi(q_1,\; q_2)$  is given by eq. (6), and $r_1$ and $r_2$ by eq. (5).

A simple calculus leads to the Plavec's formula:
\begin{equation}
\tan ^2\phi=\frac{2({q_1}_{L_1})^{-3}+2q(1-{q_1}_{L_1})^{-3}+(1+q)}{({q_1}_{L_1})^{-3}+q(1-{q_1}_{L_1})^{-3}-(1+q)}\;,
\end{equation}
where $L_1({q_1}_{L_1}\;,0)$ is the first Lagrangian point into the physical plane.

In the parametric plane we have:
\begin{equation}
\tan ^2\Phi=-\frac{\frac{\partial ^2 \Psi}{\partial Q_1^2}}{\frac{\partial ^2 \Psi}{\partial Q_2^2}},
\end{equation}
where $\Psi(Q_1,\; Q_2)$  is given by eq. (20), and $r_1$ and $r_2$ by eq. (21). A simple calculus give us:
\begin{eqnarray}
\tan ^2\Phi &=&\frac{(n+1)({Q_1}_{L_1}^n)^{-3}+q(n+1)(1-{Q_1}_{L_1}^n)^{-3}+(1+q)(2n-1)}{({Q_1}_{L_1}^n)^{-3}+q(1-{Q_1}_{L_1}^n)^{-3}-(1+q)+(n-1)q{Q_1}_{L_1}^{-n}[(1-{Q_1}_{L_1}^n)^{-3}-1]}+\nonumber\\
&+&\frac{q(n-1){Q_1}_{L_1}^{-n}[(1-{Q_1}_{L_1}^n)^{-3}-1]}{({Q_1}_{L_1}^n)^{-3}+q(1-{Q_1}_{L_1}^n)^{-3}-(1+q)+(n-1)q{Q_1}_{L_1}^{-n}[(1-{Q_1}_{L_1}^n)^{-3}-1]}
\end{eqnarray}
For $n=1$ the eq. (31) become eq. (29); this result is normal, because for $n=1$ the parametric plane coincides with the physical plane.

\section{Conclusion}

Using the Levi-Civita generalized method for regularization of
equations of motion (7)-(8), it became of interest to analyze the
topological properties of points in the parametric plane, which
correspond to equilibrium points and to curves of zero-velocity in
the physical plane. By consequence:

\begin{enumerate}
    \item As it is written in Theorem 1, to one point in the physical plane it correspond $n$ points in the parametric plane, situated in vertices of an $n$-sided regular polygon. But to the origin of the coordinate system in the physical plane it corresponds only one point, the origin of the coordinate system in the parametric plane.
    \item By consequence, we have 5 families of equilibrium points: $L_1^1,L_1^2,...,L_1^n$; $\;L_2^1,L_2^2,...,L_2^n$; ... ; $\;L_5^1,L_5^2,...,L_5^n$ in the parametric plane.
    \item Depending of the parity of $n$, we have or we haven't equilibrium points on the ordinate axis in the parametric plane; in the physical plane there aren't equilibrium points on the ordinate axis.
    \item We denote all the equilibrium points in the parametric plane as \textit{polygonal equilibrium points}. In the physical plane the equilibrium points are known as "collinear" and "triangular" equilibrium points.
    \item The equations which give us the coordinates of polygonal equilibrium points, eqs. (22)-(23) are different from those which give the coordinates of triangular and collinear equilibrium points, eqs. (18)-(19), but for $n=1$ eqs. (22)-(23) become eqs. (18)-(19).
    \item As expected, the Jacobi constant remains invariant to the transformation of Levi-Civita generalized method.
    \item Because the origin of the coordinate system in the physical plane is taken in the center of the most massive star, the asymptotic circle is transformed into an asymptotic variety in the parametric plane (see eq. (27)).
    \item In the last section we calculated the slope of the Roche variety in $L_1^1$ point (eq. (31)), and compared with the slope of the curve of zero velocity into $L_1$ in the physical plane (eq. (29)). For $n=1$ these two equations coincide.
\end{enumerate}

    We believe that all this analyze can have a theoretical importance, because it helps us to better understanding the parametric plane.

\end{document}